\documentclass[twocolumn]{aastex631}
\shorttitle{[\ion{Ne}{5}] and [\ion{Ne}{6}] in M83}
\shortauthors{Hernandez et al.}

\def\kms{\hbox{km$\,$s$^{-1}$}}

\begin{document}


\title{JWST/MIRI detection of [\ion{Ne}{5}] and [\ion{Ne}{6}] in M83: Evidence for the long sought-after AGN?}

\correspondingauthor{Svea Hernandez}
\email{sveash@stsci.edu}

\author[0000-0003-4857-8699]{Svea Hernandez}
\affiliation{AURA for ESA, Space Telescope Science Institute, 3700 San Martin Drive, Baltimore, MD 21218, USA}

\author[0000-0002-0806-168X]{Linda J. Smith}
\affiliation{Space Telescope Science Institute, 3700 San Martin Drive, Baltimore, MD 21218, USA}

\author[0000-0002-1706-7370]{Logan H. Jones}
\affiliation{Space Telescope Science Institute, 3700 San Martin Drive, Baltimore, MD 21218, USA}

\author[0000-0001-5042-3421]{Aditya Togi}
\affiliation{Department of Physics, Texas State University, 601 University Drive, San Marcos, TX 78666, USA}

\author[0000-0001-8485-0325]{Marcio B. Meléndez}
\affiliation{Space Telescope Science Institute, 3700 San Martin Drive, Baltimore, MD 21218, USA}

\author[0000-0002-2764-6069]{Valentina Abril-Melgarejo}
\affiliation{Space Telescope Science Institute, 3700 San Martin Drive, Baltimore, MD 21218, USA}

\author[0000-0002-8192-8091]{Angela Adamo}
\affiliation{The Oskar Klein Centre, Department of Astronomy, Stockholm University, AlbaNova, SE-106 91 Stockholm, Sweden}

\author[0000-0001-6794-2519]{Almudena Alonso Herrero}
\affiliation{Centro de Astrobiología (CAB) CSIC-INTA, Camino Bajo del Castillo s/n, 28692 Villanueva de la Cañada, Madrid, Spain}

\author[0000-0003-0699-6083]{Tanio Díaz-Santos}
\affiliation{Institute of Astrophysics, Foundation for Research and Technology-Hellas (FORTH), Heraklion, 70013, Greece}
\affiliation{School of Sciences, European University Cyprus, Diogenes Street, Engomi 1516, Nicosia, Cyprus}

\author[0000-0002-3365-8875]{Travis C. Fischer}
\affiliation{AURA for ESA, Space Telescope Science Institute, 3700 San Martin Drive, Baltimore, MD 21218, USA}

\author[0000-0003-0444-6897]{Santiago García-Burillo}
\affiliation{Observatorio Astronómico Nacional (OAN-IGN)-Observatorio de Madrid, Alfonso XII, 3, 28014, Madrid, Spain}

\author[0000-0002-2954-8622]{Alec S. Hirschauer}
\affiliation{Space Telescope Science Institute, 3700 San Martin Drive, Baltimore, MD 21218, USA}

\author[0000-0001-9162-2371]{Leslie K. Hunt}
\affiliation{INAF - Osservatorio Astrofisico di Arcetri, Largo E. Fermi 5, 50125, Firenze, Italy}

\author[0000-0003-4372-2006]{Bethan James}
\affiliation{AURA for ESA, Space Telescope Science Institute, 3700 San Martin Drive, Baltimore, MD 21218, USA}

\author[0000-0002-7716-6223]{Vianney Lebouteiller}
\affiliation{Université Paris-Saclay, Université Paris-Cité, CEA, CNRS, AIM, 91191, Gif-sur-Yvette, France}

\author[0000-0002-4134-864X]{Knox S. Long}
\affiliation{Space Telescope Science Institute, 3700 San Martin Drive, Baltimore, MD 21218, USA}
\affiliation{Eureka Scientific Inc., 2542 Delmar Avenue, Suite 100, Oakland, CA 94602-3017, USA}

\author[0000-0003-2589-762X]{Matilde Mingozzi}
\affiliation{ESA for AURA, Space Telescope Science Institute, 3700 San Martin Drive, Baltimore, MD21218, USA}

\author[0000-0002-9190-9986]{Lise Ramambason}
\affiliation{Institut fur Theoretische Astrophysik, Zentrum für Astronomie, Universität Heidelberg, Albert-Ueberle-Str. 2, 69120, Heidelberg, Germany}

\author[0000-0001-8353-649X]{Cristina Ramos Almeida}
\affiliation{Instituto de Astrofísica de Canarias, Calle Vía Láctea, s/n, 38205, La Laguna, Tenerife, Spain}
\affiliation{Departamento de Astrofísica, Universidad de La Laguna, 38206, La Laguna, Tenerife, Spain}



\begin{abstract}
We report the first detections of [\ion{Ne}{5}] $\lambda$14.3$\micron$ and [\ion{Ne}{6}] $\lambda$7.7 $\micron$ at high confidence (S/N $\geq$ 6) in the nuclear region of the nearby spiral galaxy M83. Emission line maps of these high ionization lines show several compact structures. Specifically, the [\ion{Ne}{6}] emission is located at 140 pc from the optical nucleus  and  appears as a point source of size $\lesssim$ 18 pc (FWHM $\lesssim$ 0.8\arcsec). We investigate the possible source of this extreme emission through comparison with photoionization models and ancillary data. We find that photoionization models of fast radiative shocks are able to reproduce the observed high excitation emission line fluxes only for the lowest preshock density available in the library, $n=$ 0.01 cm$^{-3}$. 
Additionally, tailored active galactic nuclei (AGN) photoionization models assuming a two-zone structure are compatible with the observed high ionization fluxes. Our simple AGN model shows that the emission at the location of the [\ion{Ne}{6}] source can be the result of a cloud being ionized by the radiation cone of an AGN. We stress, however, that to definitively confirm an AGN as the main source of the observed emission, more complex modeling accounting for different geometries is required. Previously known as a purely starburst system, these new findings of the nuclear region of M83 will require a reassessment of its nature and of objects similar to it, particularly now that we have access to the unparalleled infrared sensitivity and spatial resolution of the James Webb Space Telescope.

\end{abstract}



\section{Introduction}\label{sec:intro} 
Tracers of the interstellar medium (ISM), specifically the ionized gas component, are a powerful tool to characterize the physical and chemical conditions of the gas across galaxies. Fine-structure infrared (IR) lines such as [\ion{Ne}{2}] $\lambda$12.8$\micron$, [\ion{Ne}{3}] $\lambda$15.6$\micron$, [\ion{Fe}{2}] $\lambda\lambda$17.9, 26.0$\micron$, and [\ion{S}{3}] $\lambda$18.7$\micron$, allow us to access information on the mechanisms shaping the conditions of the ISM, providing a complex characterization of the gas \citep[e.g., metal content, ionization radiation field, density and temperature;][]{bal81, ver03, fer21, ric22}. The fact that IR lines are less affected by dust attenuation and extinction than their optical counterparts, makes them ideal for studying dusty environments such as star-forming regions and/or active galactic nuclei. \par

Overall, multi-wavelength emission (e.g., optical, IR) from high ionization lines such as \ion{He}{2} $\lambda$4686\AA, [\ion{O}{4}] $\lambda$25.9$\micron$ and [\ion{Fe}{5}] $\lambda$4227\AA is known to be present in star-forming galaxies (SFGs), and can serve as a probe of high-UV ionizing radiation given their photon energies \citep[e.g. $\geq$ 54 eV;][]{pak86, izo00, fri01, izo01}. Studies have reported emission from the high-ionization line [\ion{Ne}{5}] $\lambda$3426\AA\: under limited conditions, primarily in low-metallicity environments \citep[12+log(O/H) $\lesssim$ 7.88;][]{izo04, izo21}. [\ion{Ne}{5}] requires extreme-UV radiation (ionization potential, IP, of 97 eV), and it is more typically associated with the presence of active galactic nuclei \citep[AGN;][]{stu02, per22, arm23,spo22,alv23}. Although the origin of this extreme radiation in star-forming environments is still unclear, proposed mechanisms include high-mass X-ray binaries \citep[HMXBs;][]{sch19}, Wolf–Rayet (WR) stars \citep{sch96}, and fast radiative shocks \citep{dop96}. We note, however, that even to this day, photoionization models continue to encounter challenges hinting that such sources are indeed inefficient producers of the photons necessary to power these lines \citep[e.g., \ion{He}{2}, {[\ion{Ne}{5}]}; ][]{keh18, sen20, izo21}. \par
The nearby \citep[4.6 Mpc;][]{sah06} grand-design barred spiral galaxy, M83, is known to host a nuclear starburst, making it an ideal target to study star-forming conditions thoroughly. A study by \citet{fah08} reported gas spiraling from the bar into the nuclear regions of this galaxy fueling the starburst. Evidence of an age gradient in the starburst ring from both spectroscopic and imaging studies also supports the bar-driven inflow scenario \citep{har01,hou08}. Overall, after extensive studies of the nuclear region of M83, this particular galaxy has yet to show strong evidence for AGN activity, making it, as of today, a purely star-forming system \citep{tha00, sor03, kna10, del22}.

Here we report the first detections of hard ionizing radiation (photon energies $\geq$ 97 eV) in the metal-rich \citep[$\sim$1-3 Z$_\odot$][]{bre16,her21} nuclear region of M83, traced by mid-IR (MIR) emission from [\ion{Ne}{5}] (IP=$97$~eV) at 14.3$\micron$ and [\ion{Ne}{6}] (IP=$126$~eV) at 7.7$\micron$. A brief summary of the observations and data reduction is presented in Section \ref{sec:obs}. In Section \ref{sec:analysis} we detail our data analysis, and describe our results in Section \ref{sec:results}. We discuss our findings in Section \ref{sec:discussion}, and present a brief summary of our work in Section \ref{sec:conclusion}.

\section{MIRI/MRS Observations and Data Reduction}\label{sec:obs}
\begin{figure}[h]
\centering
\includegraphics[scale=0.27]{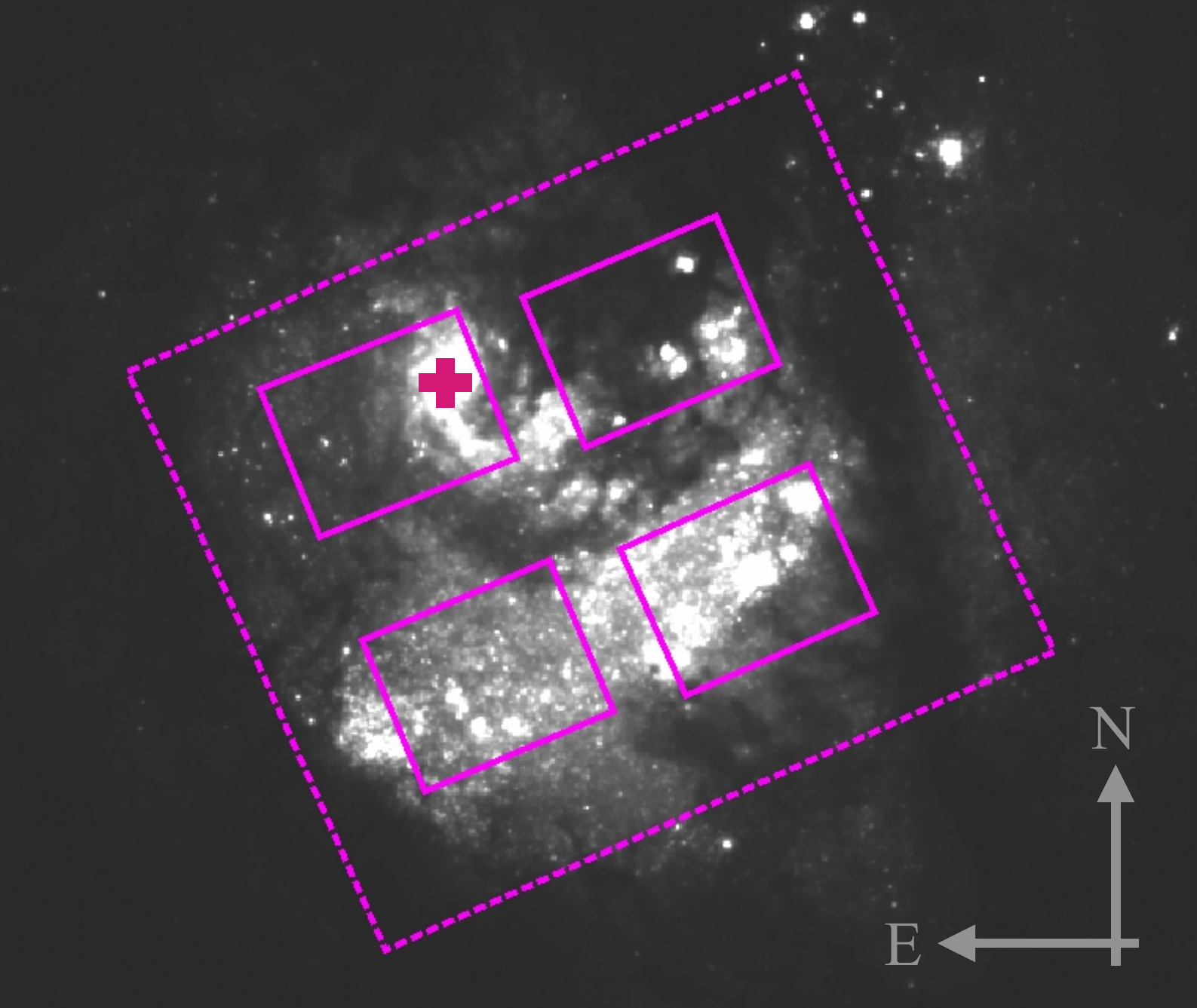}
\caption{HST/WFC3/UVIS (F814W, PID:12513) image of the core of M83. We show with magenta boxes the location of the JWST MIRI/MRS Channel 1 (solid) and Channel 4 (dashed). The cross indicates the location of the optical nucleus.} 
\label{fig:m83}
\end{figure}

The observations analyzed as part of this work were included in Cycle 1 JWST PID 02219 (PI: Hernandez); all of the JWST data used in this paper can be found in MAST:\dataset[10.17909/a61h-f081]{http://dx.doi.org/10.17909/a61h-f081}. As detailed in \citet{her23}, the data were taken with the MIRI/MRS instrument creating a 2$\times$2 mosaic, with a field of view (FOV) of 6.9\arcsec $\times$ 7.9\arcsec per pointing in channel 4, sampling the diverse environment of the nuclear starburst in M83 (Figure \ref{fig:m83}). The individual MIRI/MRS pointings strategically target the massive stellar clusters with all four MIRI/MRS channels providing contiguous wavelength coverage. \par
A complete description of the reduction steps is provided in \citet{jon24}. Briefly summarized, the data analyzed here were calibrated using the standard JWST pipeline (version 1.10.2). Given the extended nature of the starburst region of this spiral galaxy, dedicated sky observations were used to measure and correct for the thermal background. As an intermediate step, we applied fringe corrections to the 3-dimensional cubes, but we note that significant fringing persisted for a few spaxels throughout the combined cube. Lastly, during the final stage of the calibration process the combined cubes were gridded to a common scale of $0''.13$ pix$^{-1}$. We note that we focused most of our analysis on wavelengths $<$ 20 $\micron$ to avoid strongly fringed and low signal-to-noise (S/N) regions on the red end of the MIRI/MRS channel 4. We smoothed our final cube to a common resolution of $0''.788$, which is the PSF FWHM of MIRI/MRS at $\sim$20 $\micron$ \citep{law23}.

\section{Data Analysis}\label{sec:analysis}
We adopt similar methods as those detailed in \citet{jon24}. The fluxes reported in this work are corrected for attenuation using the Continuum and Feature Extraction software \citep[\texttt{CAFE}][Diaz-Santos et al., in prep.\footnote{\href{https://github.com/GOALS-survey/CAFE}{https://github.com/GOALS-survey/CAFE}}]{mar07}. Briefly, we modeled the MIR continuum as the sum of reprocessed dust and starburst continua. The available stellar population models included single stellar population (SSP) templates of ages 2, 10, 100 Myr from \texttt{STARBURST99} \citep{lei99}. Given the MIR wavelength coverage of MIRI/MRS, we included a combination of  cool (40 $-$ 100 K) and warm (100 $-$ 400 K) components to model the dust continua across the field of view of our observations \citep{mar07}.\par

To obtain emission line maps, we fit and subtract the continuum locally around the emission lines of interest. We then fit single and double Gaussian profiles to the continuum-subtracted spectra. \citet{jon24} noted that the [\ion{Ne}{2}] and [\ion{Ne}{3}] lines in this nuclear starburst were better modeled with two-component fits. Similar to their work, we apply a statistical $F$-test \citep{wes07} to determine if the individual emission lines are better represented by a single or a double Gaussian component. A two-component fit is only preferred if its reduced-$\chi^{2}$ is lower by a factor of 3.289 (corresponding to a significance level of 10\%) when compared to its one-component counterpart. We adopt a robust S/N threshold of $\geq$6. We note that application of the $F$-test confirmed that all of the [\ion{Ne}{5}] and [\ion{Ne}{6}] emission considered in this work is well-characterized by a single Gaussian profile. \par

\begin{figure*}[h]
\centering
\includegraphics[scale=0.27]{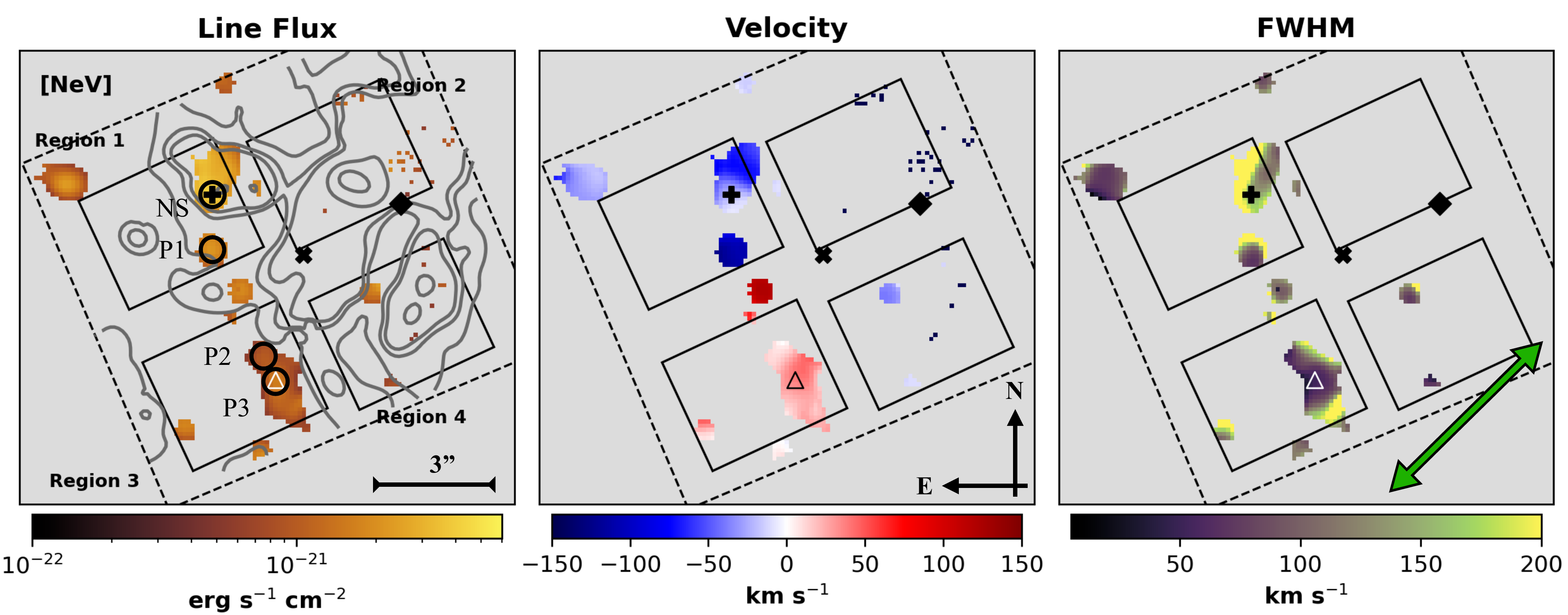}
\caption{Maps of flux (left panel), velocity offsets from systemic (middle panel), and line FWHM (right panel) for [\ion{Ne}{5}] 14.3$\mu$m. We only include spaxels with S/N $\geq$ 6. The plus symbol notes the position of the optical nucleus, the cross indicates the location of the photometric/kinematic nucleus by \citet{fah08} and \citet{kna10}, and the diamond highlights the location of the stellar kinematic center by \citet{del22}. Solid and dashed rectangles denote the channel 1 and channel 3 FoVs of our four MRS pointings. We show with a triangle the location of the source detected in [\ion{Ne}{6}] 7.7$\mu$m. We show in the left panel with black circular apertures the location of four sources (NS, P1, P2, and P3) for which we extracted contiguous spectra covering wavelengths from 4.9 to 20.5 $\micron$. In this same panel we include in grey the [\ion{Ne}{3}] emission line contours. Lastly, in the right panel we show with an arrow the orientation of the outflow reported by \citet{del22}.} 
\label{fig:nev_all}
\end{figure*}

\begin{figure*}[h]
\centering
\includegraphics[scale=0.27]{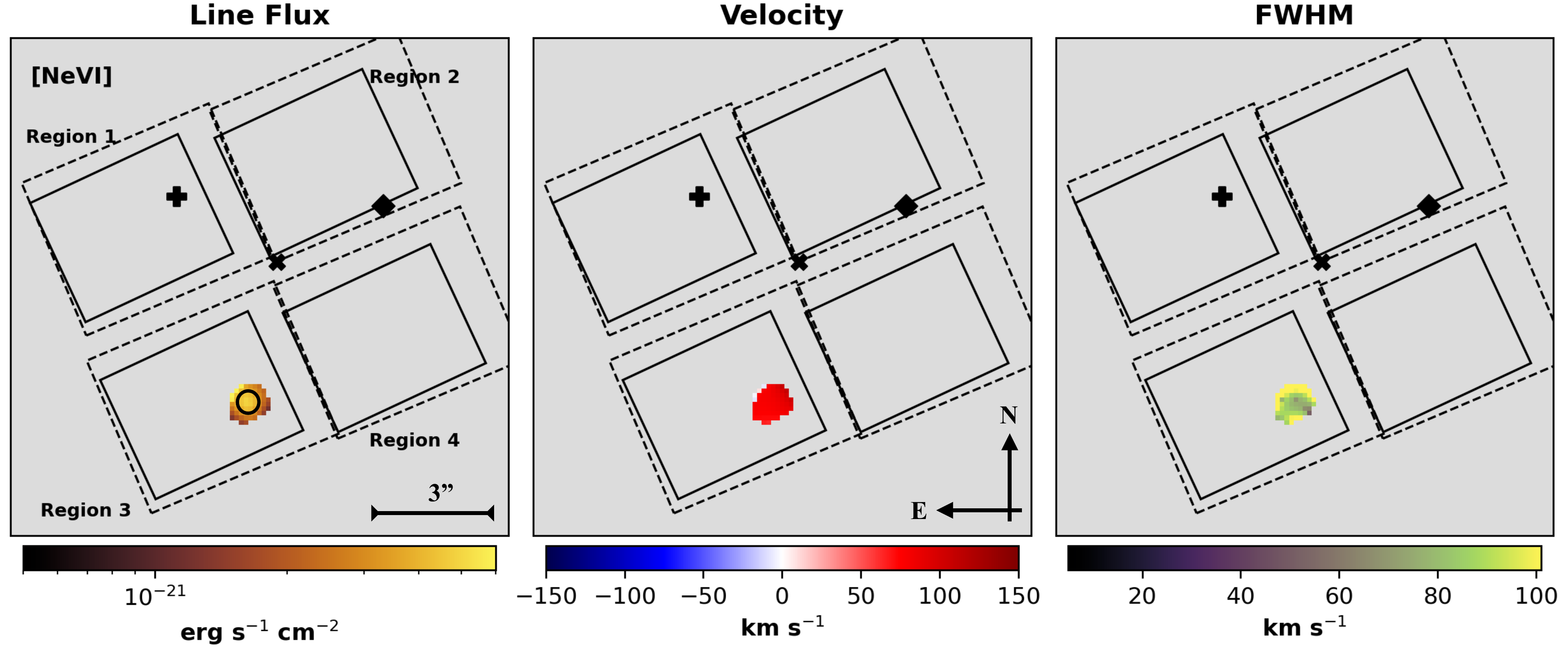}
\caption{Similar to Figure \ref{fig:nev_all}, for [\ion{Ne}{6}] 7.7$\mu$m. Solid and dashed rectangles denote the channel 1 and channel 2 FoVs of our four MRS pointings. In the left panel we show with a black circle the aperture used to extract the spectrum of the P3 source.} 
\label{fig:nevi_all}
\end{figure*}

\section{Results}\label{sec:results}
We show maps of the [\ion{Ne}{5}] 14.3$\mu$m and [\ion{Ne}{6}] 7.7$\mu$m flux, velocity, and line FWHM in Figures \ref{fig:nev_all} and \ref{fig:nevi_all}, respectively. We adopted a systemic velocity of 513 \kms \citep{all14}. Overall, the observed [\ion{Ne}{5}] emission appears to have a clumpy nature, with localized regions showing point-like structures with diameters of $\sim$ 0.7-1.1\arcsec ($\sim$ 15-24 pc)  distributed primarily throughout regions 1 and 3 (see labels in first panel in Figure \ref{fig:nev_all}). The strongest emission is observed at the location of the optical nucleus, marked with a plus symbol in Figure \ref{fig:nev_all}. It is worth noting that overall, the emission in Figure \ref{fig:nev_all} seems to be aligned in a North-to-South quasi-linear structure semi-connected to the optical nucleus. The [\ion{Ne}{5}] emission observed in region 1 tends to have negative velocities ($\gtrsim-$110 km s$^{-1}$), in contrast to the positive velocities observed outside of region 1 ($\lesssim+$120 km s$^{-1}$). This velocity field is comparable to that observed for the main component of the [\ion{Ne}{2}] $\lambda$12.8$\mu$m line reported in \citet{jon24}. In contrast to [\ion{Ne}{5}], the [\ion{Ne}{6}] emission is only detected in region 3 in an unresolved structure of $\lesssim$18 pc or FWHM $\lesssim$ 0.8\arcsec (Figure \ref{fig:nevi_all}) at a projected distance of 140 pc south of the optical nucleus or nuclear star cluster (NS). This highly ionized gas is observed at a velocity of $\sim+$80 km s$^{-1}$, higher than the velocity of the [\ion{Ne}{5}] gas at this same location ($\sim+$ 40 km s$^{-1}$). Lastly, we show in the left panel of Figure \ref{fig:nev_all} with grey  contours the observed [\ion{Ne}{3}] $\lambda$15.6$\micron$ emission, highlighting that the [\ion{Ne}{5}] and [\ion{Ne}{6}] emission detected in region 3 lies in an area of low density gas, outside of the [\ion{Ne}{3}] emission line contours.

\par
To further investigate any possible spatial differences in the spectral signatures of the [\ion{Ne}{5}] and [\ion{Ne}{6}] emission, using an aperture with radius of 0.35\arcsec\: we extracted integrated spectra for the four sources labeled as NS, P1, P2 and P3 in the first panel of Figure \ref{fig:nev_all}. We correct the extracted spectra for Galactic extinction \citep{gor21}, and intrinsic attenuation as estimated with \texttt{CAFE} (Section \ref{sec:analysis}). Figure \ref{fig:spectra_sources} shows the extracted attenuation-corrected spectra with the corresponding labels. This figure highlights the strong variations in environment, particularly visible through the strength of the broad polycyclic aromatic hydrocarbon (PAH) features and their continua. NS, which is the source at the nuclear star cluster, exhibits the strongest PAH emission as well as a steep continuum at longer wavelengths ($\gtrsim$15 $\micron$). According to \citet{mar07}, the latter is indicative of heavy obscuration. P3, on the other hand, the source farthest from the optical nucleus (as well as the only source showing [\ion{Ne}{6}] emission), displays the weakest PAH emission out of all four sources, and exhibits a much flatter continuum at longer wavelengths indicating low obscuration. \par
\begin{figure*}[h]
\centering
\includegraphics[scale=0.5]{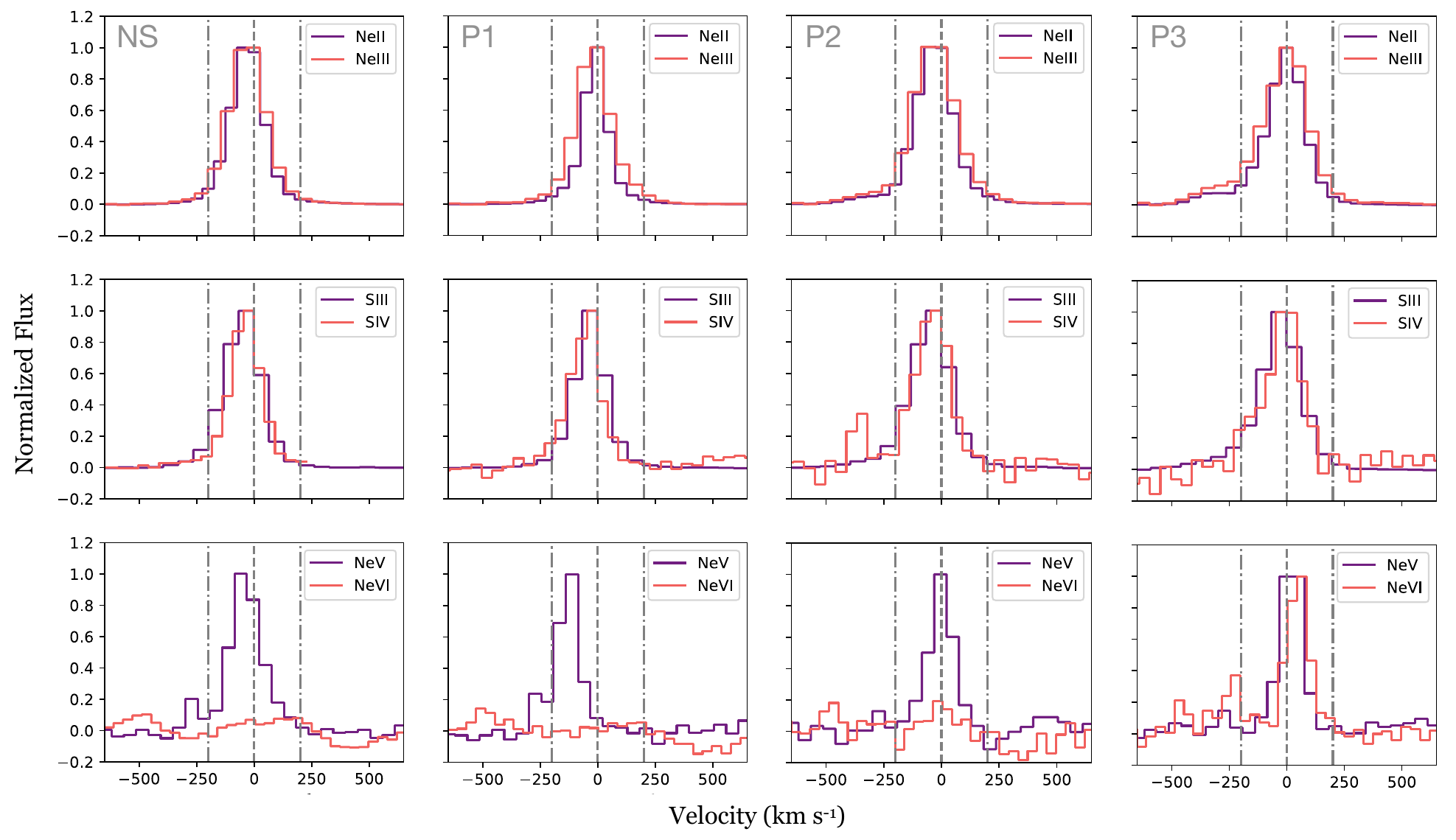} 
\caption{Normalized emission line profiles in velocity space, after correcting for systemic velocity \citep[513 km s$^{-1}$;][]{all14}. From left to right, we show the profiles of sources NS, P1, P2 and P3. The top row shows the [\ion{Ne}{2}] and [\ion{Ne}{3}] profiles. The middle row shows the [\ion{S}{3}] and [\ion{S}{4}] profiles. And lastly in the bottom row we show the [\ion{Ne}{5}] and [\ion{Ne}{6}] profiles, when present. The dashed grey line shows the 0 km s$^{-1}$ velocity, while the dotted-dashed lines indicate the $\pm$ 200 km s$^{-1}$ values. }
\label{fig:lines_comparison}
\end{figure*}

We also note that the observed properties of the extracted sources show differences in their kinematics and across different ionization species. In Figure \ref{fig:lines_comparison} we show the continuum-normalized profiles for the [\ion{Ne}{2}] $\lambda$12.8$\micron$ and [\ion{Ne}{3}] $\lambda$15.6$\micron$ lines (top row),  [\ion{S}{3}] $\lambda$18.7$\micron$ and [\ion{S}{4}] $\lambda$10.5$\micron$ (middle row), and [\ion{Ne}{5}] and [\ion{Ne}{6}], when present (bottom row). The dotted-dashed lines in each panel indicate the $\pm$ 200 km s$^{-1}$ values. Overall, the lower ionization lines (top and middle rows), show broader profiles across all sources than those observed for the [\ion{Ne}{5}] and [\ion{Ne}{6}] lines. Although not shown in Figure \ref{fig:lines_comparison}, the profiles of the [\ion{Fe}{2}] $\lambda$5.3\micron\: lines have comparable morphologies to those observed in the other lower ionization lines. Most remarkably are the differences observed in the FWHM values for P3, the region exhibiting [\ion{Ne}{6}] emission, where the FWHM values of the low ionization lines range between $\sim$120-170 km s$^{-1}$, compared to the values observed for the high ionization lines, $\sim$60-80 km s$^{-1}$. The [\ion{Ne}{5}] emission observed in P1, the source closest to the optical nucleus, shows a velocity offset of $\sim-$120 km s$^{-1}$, compared to velocities of $\sim-$50 to 0 km s$^{-1}$ for the other ions. Such a trend might hint at a scenario where an outflow driving a shock into the ambient gas produces the high ionization features we observe.
The other regions, NS, P2 and P3, do not show such a strong outflow signature. The [\ion{Ne}{6}] in P3, however, shows a slight offset at $\sim+$60 km s$^{-1}$, shifted from the [\ion{Ne}{5}] emission at the same location.\par
In the spatially-resolved study presented here we do not detect any other higher ionization MIR lines (photon energies $>$97 eV) than the reported [\ion{Ne}{5}] and [\ion{Ne}{6}] emission. We specifically searched for [\ion{Mg}{5}] $\lambda5.6\micron$ and  [\ion{Mg}{6}] $\lambda5.5\micron$ emission, as their wavelength regime is not strongly impacted by instrumental effects such as fringing. \par 

In \citet{her23}, we performed an initial analysis of the MIRI/MRS data by co-adding all the spaxels in the FoV of channel 1 (labeled as regions 1 to 4 in Figure \ref{fig:nev_all}). [\ion{Ne}{5}] was found to be present only in regions 1 and 3. In this earlier study, [\ion{O}{4}] $\lambda$25.9$\micron$ was detected due to the intrinsically higher S/N in the summed spectra. The [\ion{O}{4}]$\lambda$25.9$\micron$/[\ion{Ne}{3}]$\lambda$15.6$\micron$ ratio was found to be higher in regions 1 and 3, hinting at higher levels of ionization in these regions, compared to those observed in regions 2 and 4 \citep{wea10,mar23}.

\section{Discussion}\label{sec:discussion}
\subsection{Unusual properties of the nucleus of M83}
Observations of M83's nuclear region have previously revealed some unusual properties. \citet{tha00} obtained long-slit near-IR spectra and found two dynamical centers. The optical nucleus corresponds to a massive nuclear star cluster with a mass of $2.5 \times 10^6$ M$_\odot$ derived from the K-band luminosity and an age of 25--60 Myr adopting a distance of 3.7 Mpc \citep{vau91}. The second dynamical peak located $\approx 3''$ SW of the optical nucleus has been discussed in detail by \citet{fah08} and \citet{kna10}. The latter authors confirm that the photometric and kinematic center of M83 is offset with respect to the optical nucleus and is highly extincted. They discuss whether this photometric and kinematic center represents the true nucleus and note that attempts to locate an AGN, which could identify the actual nucleus, have been unsuccessful. To underscore the unsolved problem of the kinematic center, a relatively recent study by \citet{del22}, fitting the stellar continuum in MUSE observations, reports that the kinematic center of the stellar disk is instead located $\sim$6\arcsec\: W of the nuclear star cluster.  
In Figure ~\ref{fig:nev_all}, we indicate the positions of the potential nuclei with plus (optical nucleus or nuclear star cluster), cross (kinematic center by \citealt{fah08} and \citealt{kna10}) and diamond (MUSE stellar kinematic center by \citealt{del22}) symbols. 

The analysis of the MUSE observations also focused on the H$\alpha$ emission. \citet{del22} find evidence for a large-scale, ionized gas outflow originating from a location external to any of the previously identified centers. Note that the reported origin of this outflow is also located external to the FoV of our MIRI/MRS observations. \citet{del22} detect two cones oriented in the NW-SE direction each $20''$ in size, expanding at 100~\kms\ with a high velocity dispersion of 80--200~\kms. We show the orientation of the outflow as a green arrow in the right panel of Figure \ref{fig:nev_all}. 
They discuss the likely origin of this outflow in terms of a starburst- or an AGN-driven wind, preferring a starburst driven wind as they do not detect any shock-related emission close to the central region. Unfortunately, the FoV and the spatial scales probed by our MIRI/MRS observations make it difficult to directly connect the outflows observed by \citet{del22} with our spatially resolved study.\par

Lastly, X-ray observations using Chandra \citep{sor03, lon14}  revealed a luminous, $3.2\pm0.2 \times 10 ^{38}$~erg\,s$^{-1}$ point source coincident with the nuclear star cluster (cross symbol in Figures \ref{fig:nev_all} and \ref{fig:nevi_all}). These studies report that the source is consistent with a supermassive black hole (SMBH) with a mass of $\gtrsim$ 10$^{7}$ $M_{\odot}$, although they cannot exclude a stellar mass X-ray binary origin. Complementary to this study, through the analysis of {\it NuSTAR} observations, \citet{yuk16} identify hard X-ray emission consistent with the location of the nuclear star cluster, concluding that the emission is more likely to be associated with the integrated light from X-ray binaries than an obscured AGN. They note, however, that a SMBH accreting at a very low level cannot be ruled out.

\begin{figure*}[h]
\centering
\includegraphics[scale=0.68]{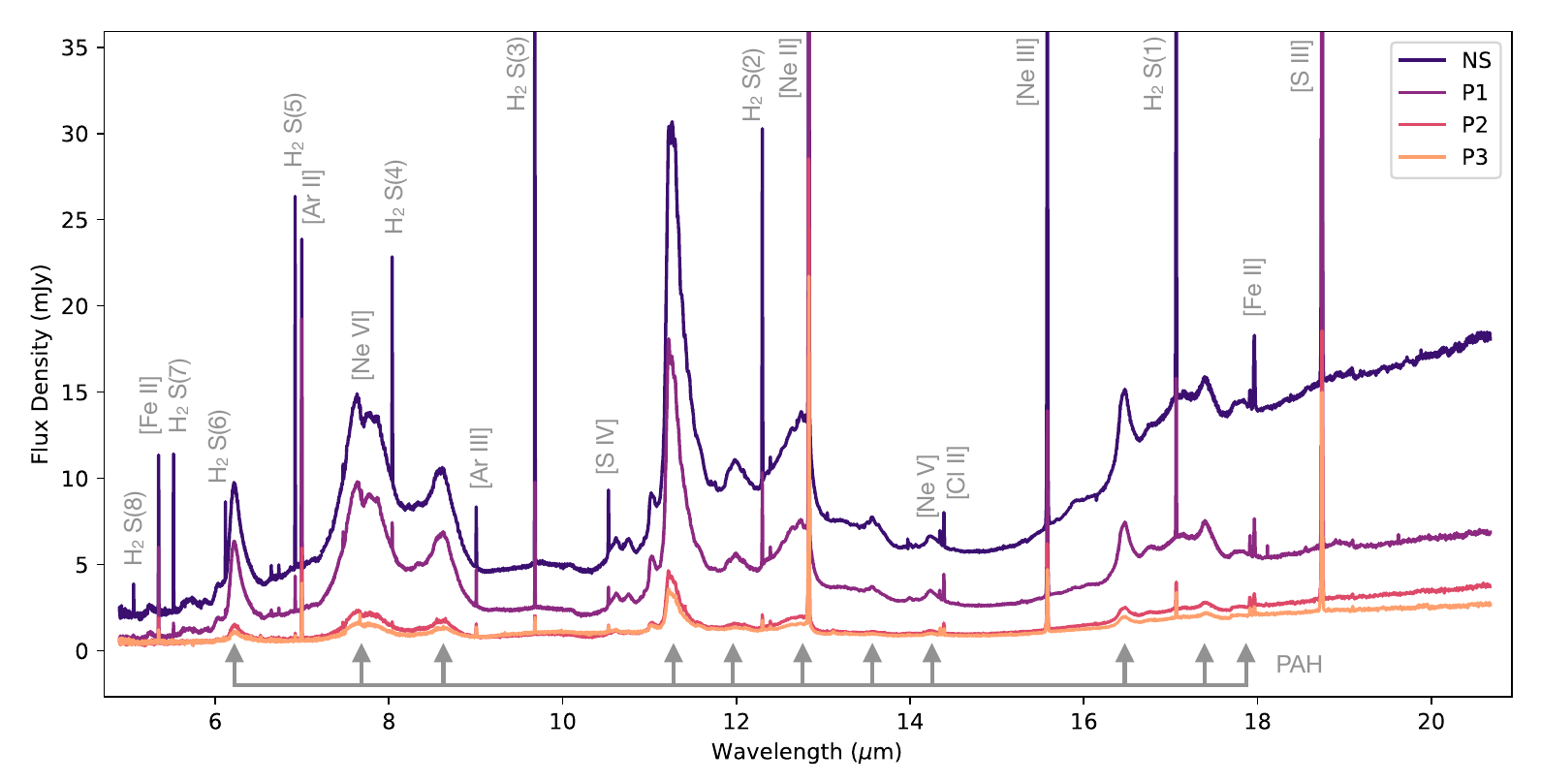}
\caption{MIR attenuation-corrected spectra of four different regions in the nucleus of M83 exhibiting [\ion{Ne}{5}] 14.3$\mu$m emission. We highlight the strongest emission lines present, including PAH features. The labels in the legend correspond to the regions highlighted with black apertures in Figure \ref{fig:nev_all}.} 
\label{fig:spectra_sources}
\end{figure*}

\begin{figure}[h]
\centering
\includegraphics[scale=0.34]{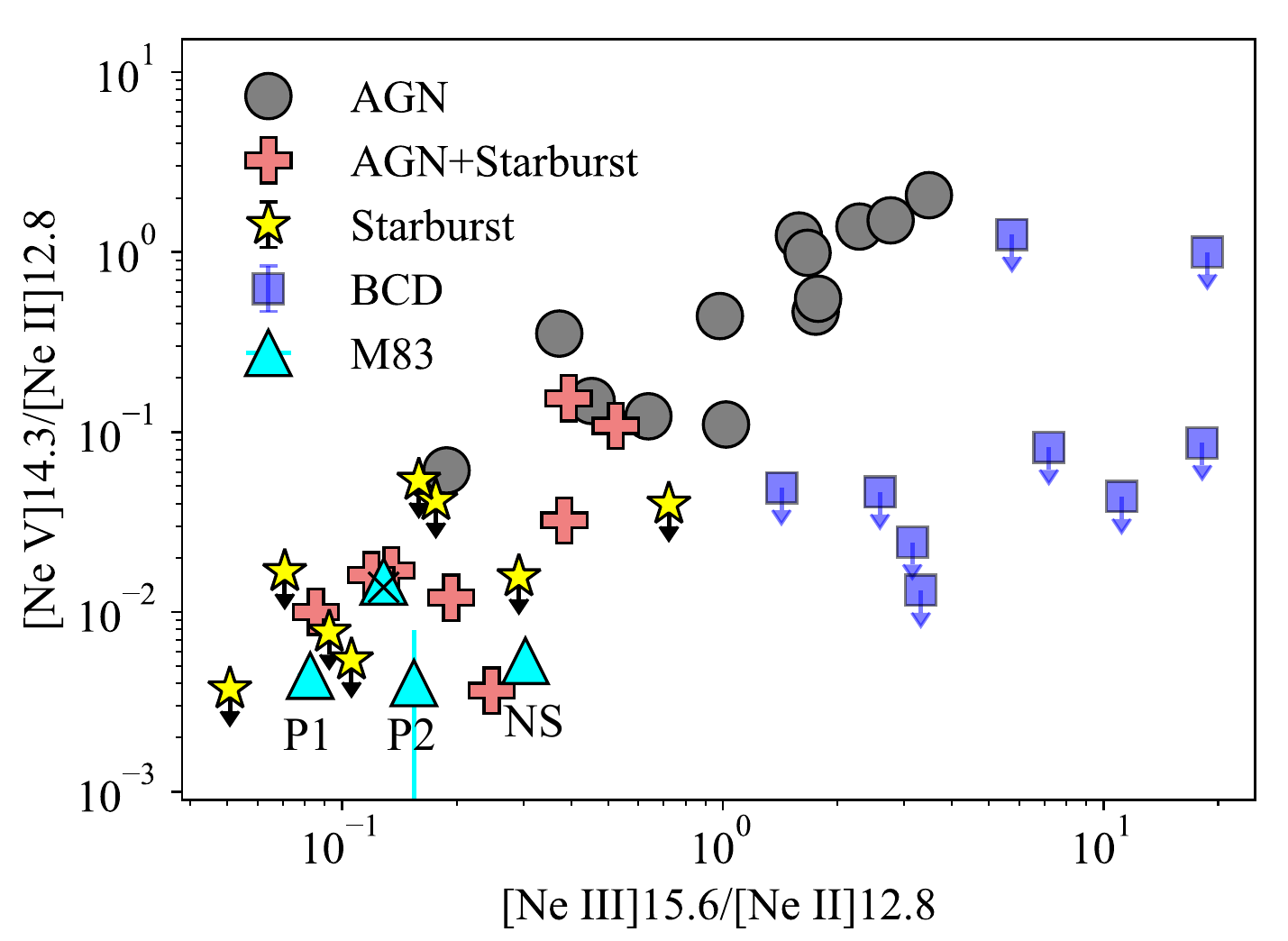}
\caption{[\ion{Ne}{5}]14.32 $\micron$/[\ion{Ne}{2}]12.81 $\micron$ as a function of the [\ion{Ne}{3}]15.56 $\micron$/[\ion{Ne}{2}]12.81 $\micron$ ratio for AGN \citep[Ramos Almeida et al. submitted; ][]{stu02, feu24, her24}, AGN+starburst \citep{sat07, sat08, her24b}, starbursts \citep{ver03}, and BCDs \citep{hao09}. The ratios observed in M83 for the three extracted regions (NS, P1, P2, and P3) are shown in cyan. We show with a cross the pointing for which we detect cospatial [\ion{Ne}{6}], P3, and include labels for the rest of the M83 pointings.} 
\label{fig:ne3ne2_ne5ne2}
\end{figure}

\subsection{Origin of the extreme emission}
[\ion{Ne}{5}] emission at $\lambda$14.3$\micron$ and/or $\lambda$24.3$\micron$ has been historically used to identify AGN activity \citep[e.g., ][]{gro06}. The power of the MIR [\ion{Ne}{5}] emission relied on the fact that it can more easily uncover the presence of AGN even when the object is obscured in the optical \citep{sat07}. One complication to this AGN diagnostic is that hard stellar radiation fields by either shocks or young stellar populations \citep[i.e. Wolf-Rayet stars and O stars;][]{con04, sch99, lei99} can in some cases also produce Ne$^{4+}$ traced by MIR [\ion{Ne}{5}] emission. In general, although [\ion{Ne}{5}] is typically linked to photoionization from AGN, in the last few decades observations have required in-depth investigations to understand under which circumstances the [\ion{Ne}{5}] emission has a component, or is fully produced, by star formation or radiative shocks. \par

Using Spitzer, studies such as those by \citet{sat07,sat08} discovered a sample of SFGs classified as such through optical diagnostics but exhibiting MIR [\ion{Ne}{5}] emission. Through comparison with photoionization models that include both components, AGN and starburst contributions, they report that the [\ion{Ne}{5}] emission observed in ``starburst'' systems can only be reproduced when including an AGN component. For some galaxies they find AGN contributions as low as 10\% of the total galaxy luminosity. This population of galaxies was previously identified as purely starburst given that the fraction of their total luminosity due to the AGN is low. In this regime, optical diagnostics are insensitive to the presence of the AGN \citep{ter95, sat08}. It is in this parameter space that MIR emission is critical to the previously-missed presence of possibly weak AGN. \par
Ratios of high-to-low excitations are used to characterize the ionizing source. More specifically, the [\ion{Ne}{5}] $\lambda$14.3$\micron$/[\ion{Ne}{2}] $\lambda$12.8$\micron$ line flux ratio has served to identify the nature of the main ionizing source in galaxies \citep{stu02,arm07, feu24}. In Figure \ref{fig:ne3ne2_ne5ne2} we show the [\ion{Ne}{5}] $\lambda$14.3$\micron$/[\ion{Ne}{2}] $\lambda$12.8$\micron$ ratio as a function of [\ion{Ne}{3}] $\lambda$15.6$\micron$/[\ion{Ne}{2}] $\lambda$12.8$\micron$ for different galaxy classifications. We show with gray circles the ratios observed in typical AGN systems as reported by \citet{stu02}, and more recently by \citet{feu24},  \citet{her24} and Ramos Almeida et al. (\textit{submitted}) using JWST. The upper limits recorded from starburst systems are shown with star symbols \citep{ver03}. Similarly, the upper limits for compact systems, such as blue compact dwarfs (BCDs), are shown in purple squares; those targets for which \citet{sat07,sat08} report a combined contribution of AGN and starburst, along with recent JWST results by \citet{her24b} for a similar system, are displayed with plus symbols. Lastly, the ratios observed in the spectra extracted from four different regions in the nucleus of M83 are shown with cyan triangles. Overall, the neon ratios for AGN systems typically have relatively high [\ion{Ne}{5}]/[\ion{Ne}{2}] ratios, but lower [\ion{Ne}{3}]/[\ion{Ne}{2}] values than those recorded for BCDs. This is an indication that the ionization state of AGN environments tends to be higher than in any other system. As expected, starburst galaxies exhibit lower [\ion{Ne}{5}]/[\ion{Ne}{2}] ratios than AGNs, but we note the continuous distribution between AGN and starburst systems which is clearly bridged by the confirmed AGN+starburst galaxies (pink plus symbols). \par
From Figure \ref{fig:ne3ne2_ne5ne2} it is evident that overall the observed neon emission in the nucleus of M83 appears to overlap with the observed upper limits in starburst galaxies. More specifically, we highlight that the neon emission detected in P3, the source exhibiting cospatial [\ion{Ne}{6}] emission (shown with a cross in Figure \ref{fig:ne3ne2_ne5ne2}) lands on the parameter space covered by the upper limits of the purely-starburst systems and the AGN+starburst galaxies.\par

\begin{figure}
\centering
\includegraphics[scale=0.62]{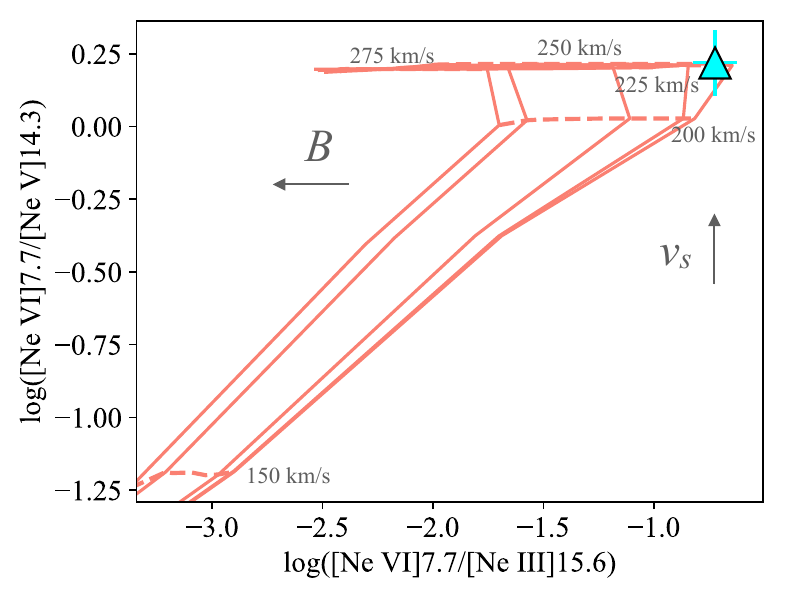}
\caption{[\ion{Ne}{6}]7.7 $\micron$/[\ion{Ne}{5}]14.3 $\micron$ vs. [\ion{Ne}{6}]7.7 $\micron$/[\ion{Ne}{3}]15.6 $\micron$ diagnostic diagram for the solar abundance, $n$= 0.01 cm$^{-3}$ shock-only models by \citet{all08}. We show with a triangle the emission line ratios of P3, the only source where we detect [\ion{Ne}{6}].} 
\label{fig:shock_model}
\end{figure}
\subsubsection{Radiative shocks}
In general, photoionization models with a starburst component alone as the main source of ionization struggle to produce the [\ion{Ne}{5}] emission observed in SFGs \citep{thu05,feu24}. A promising candidate for the high-ionization emission lines and hard radiation traced by  [\ion{Ne}{5}] is instead fast radiative shocks  \citep[e.g.,][]{izo04,thu05}. These radiative shock waves are expected from astrophysical objects or events such as outflows from young stellar systems, supernovae, and outflows from active/starburst galaxies. Under the assumption that the observed emission is originating from shocks, we now consider fast radiative shocks where the kinetic energy of such supersonic motions is dissipated through these radiative processes. In general, ionizing radiation generated by the shock fronts creates a strong radiation field of extreme UV and soft X-ray photons which translates into appreciable photoionization effects \citep{sut93, dop03}. \par

We explore the grid of fast radiative shock models by \citet{all08}. The properties of these models are defined by four physical parameters: shock velocity, $v_{s}$; preshock transverse magnetic field, $B$; preshock density, $n$; and metallicity or abundances. The only models in this library capable of producing the observed [\ion{Ne}{5}] and [\ion{Ne}{6}] emission observed in P3 are those with solar abundance and the lowest preshock density available, $n$= 0.01 cm$^{-3}$. In Figure \ref{fig:shock_model} we show the [\ion{Ne}{6}]7.7 $\micron$/[\ion{Ne}{5}]14.3 $\micron$ vs. [\ion{Ne}{6}]7.7 $\micron$/[\ion{Ne}{3}]15.6 $\micron$ diagnostic diagram along with the observed emission line fluxes for P3, the only source exhibiting [\ion{Ne}{6}] emission. According to these models, the extreme [\ion{Ne}{5}] and [\ion{Ne}{6}] emission can originate from radiative shocks with velocities of $\sim$ 225-250 km s$^{-1}$ and a low preshock transverse magnetic field of $B$= 0.001 $\mu$G. We note that although the shock models are able to reproduce the observed emission line ratios, at the lowest shock velocities, $\sim$100 km s$^{-1}$, the models are incapable of producing any high ionization emission, i.e., [\ion{Ne}{6}]. Overall, energetic shocks would efficiently destroy PAH molecules in their vicinity, which is what is observed when comparing the relative strength of the PAH features in P3 against those in the other three regions \citep[Figure \ref{fig:spectra_sources};][]{per22}. \par
To get a general sense of the observed densities, or post-shock densities, at the location of the [\ion{Ne}{6}] source, we estimate an upper limit flux of $4.8\times10^{-20}$ $W\: m^{-2}$ for the [\ion{Ne}{5}] $\lambda24.2\micron$ line and use it along with the measured flux for the [\ion{Ne}{5}] $\lambda14.3\micron$ line to infer a lower limit on the density, \textit{n}. From this ratio we estimate a post-shock density of $>$1000 cm$^{-3}$, higher than the pre-shock density predicted by the models by more than five orders of magnitude. Although high densities are expected behind the shock front due to radiative cooling and gas compression \citep{all08, ala19}, we note that the predicted pre-shock density that matches our observed [\ion{Ne}{6}]/[\ion{Ne}{5}] ratio is extremely low at 0.01  cm$^{-3}$ and at the limit of the models calculated by \citet{all08}. 
\par

Since we have identified radiative shock models capable of reproducing the emission line ratio observed at P3, some possible sources producing these shocks are supernovae. Indeed, the nuclear region of M83 has been reported to host an extensive population of supernova remnants \citep[SNRs; ][]{rus20}. We compiled a list of confirmed SNRs from different catalogues \citep[SNRs; ][]{dop10, bla12, bla14, lon22} and cross-correlated the location of these sources with those from our [\ion{Ne}{5}] and [\ion{Ne}{6}] sources. Unfortunately, none of the SNRs detected so far coincide with the location of the sources studied here. We also note that region 3 in our observations (Figure \ref{fig:nev_all}) is particularly devoid of SNRs.

Lastly, we highlight that this work assumes that P3, the [\ion{Ne}{6}] source, is part of the same structure as the rest of the surrounding emission, i.e., co-spatial with the lower ionization gas ([\ion{Ne}{2}], [\ion{Ne}{3}]). This assumption excludes a scenario where the [\ion{Ne}{6}] source is a compact nebula embedded in the extended diffuse ISM. If this alternative scenario is instead adopted, the lower-ionization fluxes (e.g., [\ion{Ne}{3}]) would need to be corrected by removing the contributions of the diffuse emission. Doing this would then result in lower fluxes for [\ion{Ne}{3}], shifting the P3 point in Figure \ref{fig:shock_model} to higher values of [\ion{Ne}{6}]/[\ion{Ne}{3}], and outside of the parameter space covered by the most extreme shock models. Under these considerations shocks are unlikely to be the source of the extreme radiation. \par

\begin{figure*}
\centering
\includegraphics[scale=0.27]{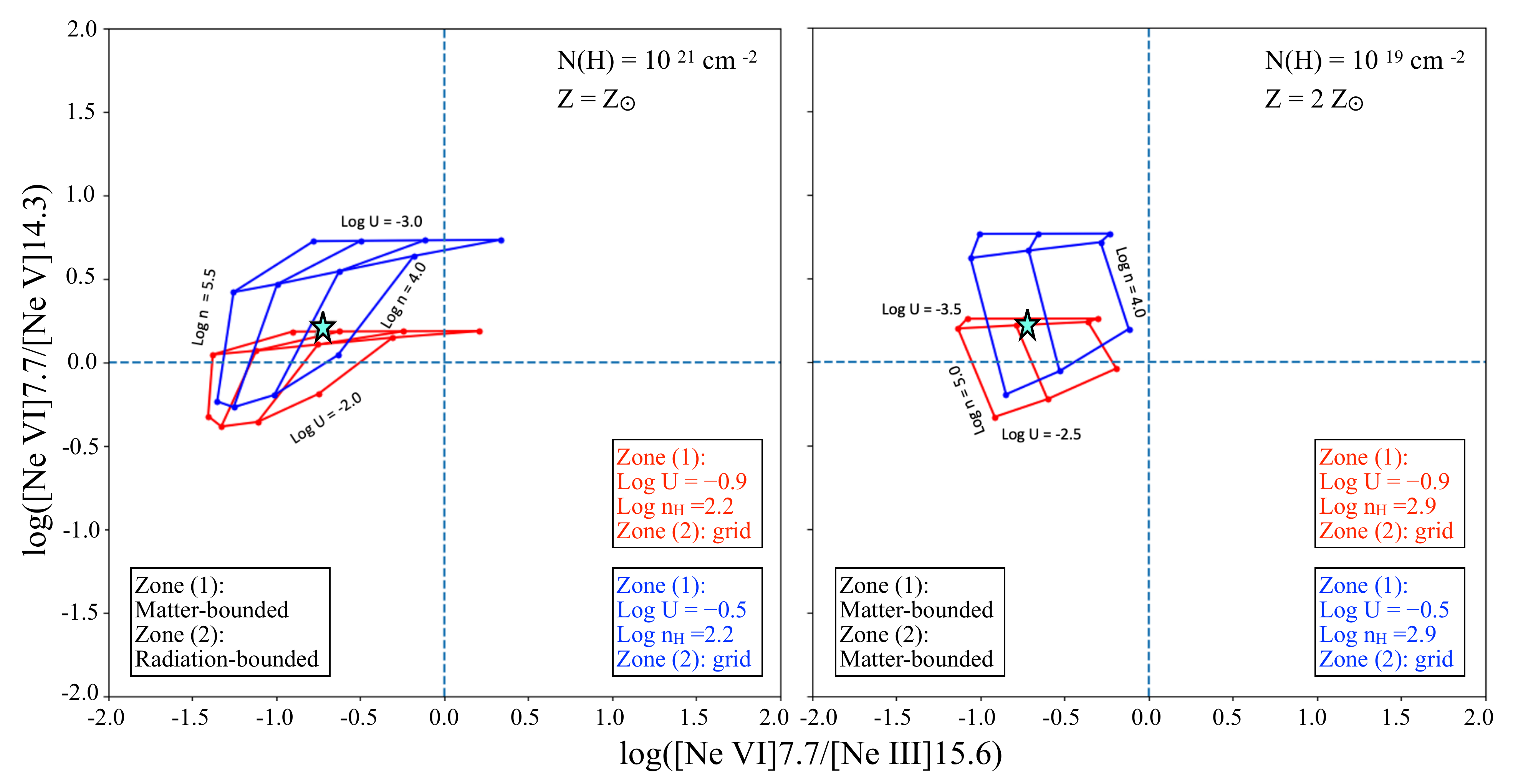}
\caption{Two-zone AGN photoionization models. \textit{Left:} Models assuming a hydrogen column density of 10$^{21}$ cm$^{-2}$ and solar abundance, Z$_{\odot}$. In red we show the MIR diagnostics predicted adopting an ionization parameter log $U=-$0.9 and density of log $n_{\rm H}=$2.2 for the matter-bounded zone (1), the grid shows the varying parameters for the zone (2) radiation-bounded component. In blue we show the predicted grid for a model adopting an ionization parameter log $U=-$0.5 and density of log $n_{\rm H}=$2.2 for the matter-bounded zone (1), the grid highlights the varying parameters for the zone (2) radiation-bounded component. \textit{Right:} Models assuming a hydrogen column density of 10$^{19}$ cm$^{-2}$ and twice solar abundance, 2 Z$_{\odot}$, properties aligned with those observed in the nucleus of M83 \citep{her21}. Given the relatively low column density, both zones are matter-bounded. In red we show the ratios predicted adopting an ionization parameter log $U=-$0.9 and density of log $n_{\rm H}=$2.9 for the matter-bounded zone (1), the grid shows the varying parameters for the matter-bounded zone (2). In blue we show the predicted grid for a model adopting an ionization parameter log $U=-$0.5 and density of log $n_{\rm H}=$2.9 for the matter-bounded zone (1). In both panels, the observed emission line ratios in M83 are shown with a cyan star.} 
\label{fig:agn_model}
\end{figure*}
\subsubsection{AGN}
Given that ionization energies high enough to generate [\ion{Ne}{5}] and [\ion{Ne}{6}] emission is commonly expected to be produced by AGN \citep{mel08, per10, mor21}, we turn to photoionization models to explain the observed emission assuming an AGN as the main ionization source. The photoionization models relied on the version c23.01 of the spectral synthesis code \texttt{CLOUDY} \citep{cha23, gun23}. Similar to the work by \citet{mel11}, we generate AGN photoionization models assuming a broken power-law continuum with an adopted EUV-soft X-Ray slope of $\alpha=$ 1.9. This slope represents the median value in Seyfert 1 galaxies based on their observed high ionization MIR lines. We adopt a two-zone approximation to match all of the observed emission line fluxes listed in Table \ref{table:pah}. This initial two-zone approach assumes a combination of (1) a matter-bounded component optically thin to the ionizing radiation where the high ionization lines [\ion{Ne}{5}] and [\ion{Ne}{6}] are emitted efficiently, and (2) a radiation-bounded component optically thick to ionizing radiation, which maximizes lower ionization emission, e.g., [\ion{S}{4}], [\ion{Ne}{3}] and [\ion{Ne}{2}]. \par
For this initial model we adopt a typical AGN hydrogen column density of 10$^{21}$ cm$^{-2}$ \citep{per10,mel11}, and solar abundances. In the left panel of Figure \ref{fig:agn_model} we show two tailored photoionization models assuming different physical properties for zone (1) and zone (2). In red we show the MIR diagnostics predicted adopting an ionization parameter log $U=-$0.9 and density of log $n_{\rm H}=$2.2 for the matter-bounded zone (1), with the grid showing the varying parameters for the zone (2) radiation-bounded component. In blue we show a separate model adopting an ionization parameter log $U=-$0.5 and density of log $n_{\rm H}=$2.2 for the matter-bounded zone (1), with the grid highlighting the varying parameters for the zone (2) radiation-bounded component. The observed emission line ratio in M83, specifically for P3, is shown with a cyan star symbol. They are reproduced by the blue grid in Figure \ref{fig:agn_model}, which assumes log $U=-$0.5 and log $n_{\rm H}=$2.2 for the zone (1) component, and log $U=-$2.9 and log $n_{\rm H}=$4.6 for the zone (2) component. We highlight that these model assumptions and the optimized solution capable of reproducing our observed fluxes is consistent with previous modeling of the narrow-line region (NLR) in typical AGNs \citep{mel11,mel14}.\par

  \begin{table*}
\caption{Line fluxes and physical properties of high ionization sources}
\label{table:pah}
\tabletypesize{\footnotesize}
\centering 
\begin{tabular}{crrrrr}
\hline \hline
\multicolumn{2}{c}{Line} &  \multicolumn{1}{c}{NS}&  \multicolumn{1}{c}{P1} &  \multicolumn{1}{c}{P2} &  \multicolumn{1}{c}{P3}  \\
\multicolumn{2}{c}{Rest Wavelength} & \multicolumn{4}{c}{Flux} \\
\multicolumn{2}{c}{($\mu$m)} & \multicolumn{4}{c}{(10$^{-20}$ $W\; m^{-2}$)} \\
\hline
{[\ion{S}{4}]} & 10.5 & 62.9$\pm$1.8 &7.5$\pm$1.5 & 9.4$\pm$0.7  & 6.0$\pm$0.6  \\
{[\ion{Ne}{2}]} & 12.8 & 996.0$\pm$32.6 &550.0$\pm$21.4 & 363.0$\pm$10.9  & 251.0$\pm$8.3  \\
{[\ion{Ne}{5}]} & 14.3 & 5.32$\pm$1.4 &2.4$\pm$0.6 & 1.5$\pm$1.4  & 3.7$\pm$0.5  \\
{[\ion{Ne}{3}]} & 15.6 & 302.0$\pm$4.9 &45.4$\pm$2.8 & 56.1$\pm$1.4  & 32.3$\pm$1.7  \\
{[\ion{S}{3}]} & 18.7 & 314.0$\pm$30.0 &230.0$\pm$14.9 & 155.0$\pm$5.1  & 141.0$\pm$2.9  \\
{[\ion{Ne}{6}]} & 7.7 & $-$ &$-$ & $-$  & 6.1$\pm$1.4  \\

\hline 
\multicolumn{2}{c}{Distance from NS (pc)} & $-$ & 41 & 122 & 140 \\
\hline
\end{tabular}
\end{table*}

We note that although the adopted column density of 10$^{21}$ cm$^{-2}$ and solar abundances are typical of AGN environments \citep{mel11}, these values differ from what has been reported for the nuclear region of M83 \citep{her21}. When assuming instead a lower hydrogen column density of the order of 10$^{19}$ cm$^{-2}$, and a 2 $\times$ Z$_{\odot}$ abundance, our two-zone photoionization models predict a system with two matter-bounded components (right panel in Figure \ref{fig:agn_model}). Under these conditions, more accurately describing the observed environment in the nuclear starburst in M83, we are able to reproduce the observed emission assuming a matter-bounded zone (1) with log $U=-$0.9 and log $n_{\rm H}=$2.9 and matter-bounded zone (2) with log $U=-$3.0 and a higher hydrogen number density of log $n_{\rm H}=$4.5. \par
Overall, our work shows that the observed high ionization fluxes in the nuclear starburst in M83 are compatible with the emission from a cloud ionized by the radiation cone of an AGN. Both our MIRI/MRS observations and the AGN modeling described in this section align well with the previously proposed scenario of an SMBH accreting at a very low level \citep{sor03, lon14, yuk16}. We note, however, that although our modeling can reproduce the observed fluxes assuming an AGN as the ionizing source, more complex simulations taking into account possible geometries are required before an AGN source is confirmed as the origin of the high ionization emission. These tailored models should be able to reproduce simultaneously the emission from the different [\ion{Ne}{5}] sources (e.g., NS, P1, P2).\par

\subsubsection{Other sources?}
Hot massive stars are known to be the sources of ionizing radiation in star-forming environments. A few studies, however, have argued that these same young massive stars might also be responsible for the hard ionizing radiation capable of producing \ion{He}{2} and [\ion{Ne}{5}], but only in low-metallicity galaxies \citep[][]{sch99, koj21, min25}. Even though models which include these massive stars continue to struggle in reproducing the observed high ionization emission in these metal-poor systems \citep{thu05,izo21}, the observed conditions in M83, specifically at the nucleus, differ considerably from these environments (high metallicity, age of nuclear star cluster of 25-60 Myr, etc.), making these sources the unlikely origin of the detected  [\ion{Ne}{5}] and [\ion{Ne}{6}] emission. We also note that the stellar atmosphere models used by \citet{sch99} omit stellar wind line blanketing so the emergent fluxes below 228~\AA\ are over-predicted. Models containing wind blanketing \citep[e.g., ][]{smith02} show that massive star winds are opaque above 54~eV at the high metallicities observed at the center of M83, excluding these massive stars as the source of the observed high ionization.
\par

Lastly, since the discovery of high ionization  emission (e.g., \ion{He}{2}) in star-forming systems, additional sources that can efficiently emit energetic photons beyond what stellar sources can produce have been suggested. Some of these sources include X-ray binaries \citep[e.g., ][]{sch19}. Given that the nuclear region of M83 has been extensively studied in X-rays, we attempted to cross match the location of existing X-ray sources from the catalog by \citet{lon14} with the observed MIR [\ion{Ne}{5}] and [\ion{Ne}{6}] emission. With the exception of the X-ray source and emission located in the nuclear star cluster (optical nucleus), no other X-ray emission coincided with the high ionization emission in our MIR maps, including the [\ion{Ne}{6}] source in P3. Although X-ray emission has not been reported at the location of the [\ion{Ne}{6}] source, or any of the other sources (e.g., P1, P2), we note that faint high-mass X-ray binary (HMXB) systems cannot be ruled out as the origin of this extreme emission, particularly objects such as accreting HMXB at low luminosities, which may be as faint as 10$^{34}$ erg s$^{-1}$ \citep[e.g., ][]{rut07}. More specifically we note that the compact nature of the [\ion{Ne}{6}] source would be compatible with an in-situ source.

\section{Conclusions}\label{sec:conclusion}
We report the first detections of highly ionized [\ion{Ne}{6}] 7.7 $\micron$ and [\ion{Ne}{5}] 14.3 $\micron$ in the nuclear region of M83. The high-ionization [\ion{Ne}{6}] emission constitutes the first account in what was historically known as a  purely-starbursting system, i.e., SAB(s). We note that although our work confirms the presence of this high ionization emission at high confidence (S/N$\geq$6), the observations at hand do not show robust evidence of spatially-resolved emission from [\ion{O}{4}] 25.9 $\micron$ and/or [\ion{Ne}{5}] 24.3 $\micron$, as these lines fall in the lower-sensitivity wavelength regime, which is also strongly impacted by fringing. \par
We investigate possible mechanisms responsible for the observed high ionization emission. We identify fast radiative shock models capable of reproducing the observed MIR emission with shock velocities of $\sim$225-250 km s$^{-1}$ and a remarkably low preshock density of $n=$ 0.01 cm$^{-3}$. Additionally, our comparison with simple tailored two-zone models confirm AGN as a possible candidate responsible for the high ionization radiation generating the measured [\ion{Ne}{5}] and [\ion{Ne}{6}] emission. We stress that to definitively confirm AGN as the main source of the ionization observed in the nucleus of M83, more complex modeling accounting for different geometries is required. \par 
Lastly, we tried to cross match the location of existing X-ray sources from the catalog by \citet{lon14} with the observed MIR [\ion{Ne}{5}] and [\ion{Ne}{6}] emission and found that with the exception of the X-ray source located at the optical nucleus (NS), no other X-ray emission coincided with the high ionization emission in our MIR maps. We note, however, that a faint or embedded HMXB source cannot be ruled out as the origin of the [\ion{Ne}{6}] emission. Furthermore, given the relatively extended nature of the observed  [\ion{Ne}{5}] emission, a feasible scenario could be a combination of sources (shocks, HMXB, AGN) producing the high ionization traced by these MIR lines.\par
We highlight that this unique detection was only possible due to the unparalleled MIR sensitivity and spatial resolution afforded by JWST/MIRI MRS. In this new observational era, studies such as the one presented here begin to hint at a scenario where sources previously known as purely starburst systems, similar to M83, may require a reassessment of their nature.


\begin{acknowledgments}
We thank the referee for their careful review and constructive comments on this manuscript. SH, TCF, BJ, and MM are thankful for support from the European Space Agency (ESA). AAH acknowledges support from grant PID2021-124665NB-I00 funded by MCIN/AEI/10.13039/501100011033 and by ERDF A way of making Europe. TDS acknowledges the research project was supported by the Hellenic Foundation for Research and Innovation (HFRI) under the ``2nd Call for HFRI Research Projects to support Faculty Members \& Researchers" (Project Number: 03382). LR gratefully acknowledges funding from the DFG through an Emmy Noether Research Group (grant number CH2137/1-1). This work is based on observations made with the NASA/ESA/CSA JWST. Support for program JWST-GO-02291 was provided by NASA through a grant from the Space Telescope Science Institute, which is operated by the Associations of Universities for Research in Astronomy, Incorporated, under NASA contract NAS 5-26555. The data were obtained from the Mikulski Archive for Space Telescopes at the Space Telescope Science Institute, which is operated by the Association of Universities for Research in Astronomy, Inc., under NASA contract NAS 5-03127 for JWST. These observations are associated with program \#02219.
\end{acknowledgments}

\bibliography{sample631}{}

\begin{thebibliography}{}
\expandafter\ifx\csname natexlab\endcsname\relax\def\natexlab#1{#1}\fi
\providecommand{\url}[1]{\href{#1}{#1}}
\providecommand{\dodoi}[1]{doi:~\href{http://doi.org/#1}{\nolinkurl{#1}}}
\providecommand{\doeprint}[1]{\href{http://ascl.net/#1}{\nolinkurl{http://ascl.net/#1}}}
\providecommand{\doarXiv}[1]{\href{https://arxiv.org/abs/#1}{\nolinkurl{https://arxiv.org/abs/#1}}}

\bibitem[{{Alarie} \& {Morisset}(2019)}]{ala19}
{Alarie}, A., \& {Morisset}, C. 2019, \rmxaa, 55, 377,
  \dodoi{10.22201/ia.01851101p.2019.55.02.21}

\bibitem[{{Allen} {et~al.}(2008){Allen}, {Groves}, {Dopita}, {Sutherland}, \&
  {Kewley}}]{all08}
{Allen}, M.~G., {Groves}, B.~A., {Dopita}, M.~A., {Sutherland}, R.~S., \&
  {Kewley}, L.~J. 2008, \apjs, 178, 20, \dodoi{10.1086/589652}

\bibitem[{{Allison} {et~al.}(2014){Allison}, {Sadler}, \& {Meekin}}]{all14}
{Allison}, J.~R., {Sadler}, E.~M., \& {Meekin}, A.~M. 2014, \mnras, 440, 696,
  \dodoi{10.1093/mnras/stu289}

\bibitem[{{{\'A}lvarez-M{\'a}rquez} {et~al.}(2023){{\'A}lvarez-M{\'a}rquez},
  {Labiano}, {Guillard}, {Dicken}, {Argyriou}, {Patapis}, {Law}, {Kavanagh},
  {Larson}, {Gasman}, {Mueller}, {Alberts}, {Brandl}, {Colina},
  {Garc{\'\i}a-Mar{\'\i}n}, {Jones}, {Noriega-Crespo}, {Shivaei}, {Temim}, \&
  {Wright}}]{alv23}
{{\'A}lvarez-M{\'a}rquez}, J., {Labiano}, A., {Guillard}, P., {et~al.} 2023,
  \aap, 672, A108, \dodoi{10.1051/0004-6361/202244880}

\bibitem[{{Armus} {et~al.}(2007){Armus}, {Charmandaris}, {Bernard-Salas},
  {Spoon}, {Marshall}, {Higdon}, {Desai}, {Teplitz}, {Hao}, {Devost}, {Brandl},
  {Wu}, {Sloan}, {Soifer}, {Houck}, \& {Herter}}]{arm07}
{Armus}, L., {Charmandaris}, V., {Bernard-Salas}, J., {et~al.} 2007, \apj, 656,
  148, \dodoi{10.1086/510107}

\bibitem[{{Armus} {et~al.}(2023){Armus}, {Lai}, {U}, {Larson}, {Diaz-Santos},
  {Evans}, {Malkan}, {Rich}, {Medling}, {Law}, {Inami}, {Muller-Sanchez},
  {Charmandaris}, {van der Werf}, {Stierwalt}, {Linden}, {Privon},
  {Barcos-Mu{\~n}oz}, {Hayward}, {Song}, {Appleton}, {Aalto}, {Bohn},
  {B{\"o}ker}, {Brown}, {Finnerty}, {Howell}, {Iwasawa}, {Kemper}, {Marshall},
  {Mazzarella}, {McKinney}, {Murphy}, {Sanders}, \& {Surace}}]{arm23}
{Armus}, L., {Lai}, T., {U}, V., {et~al.} 2023, \apjl, 942, L37,
  \dodoi{10.3847/2041-8213/acac66}

\bibitem[{{Baldwin} {et~al.}(1981){Baldwin}, {Phillips}, \&
  {Terlevich}}]{bal81}
{Baldwin}, J.~A., {Phillips}, M.~M., \& {Terlevich}, R. 1981, \pasp, 93, 5,
  \dodoi{10.1086/130766}

\bibitem[{{Blair} {et~al.}(2012){Blair}, {Winkler}, \& {Long}}]{bla12}
{Blair}, W.~P., {Winkler}, P.~F., \& {Long}, K.~S. 2012, \apjs, 203, 8,
  \dodoi{10.1088/0067-0049/203/1/8}

\bibitem[{{Blair} {et~al.}(2014){Blair}, {Chandar}, {Dopita}, {Ghavamian},
  {Hammer}, {Kuntz}, {Long}, {Soria}, {Whitmore}, \& {Winkler}}]{bla14}
{Blair}, W.~P., {Chandar}, R., {Dopita}, M.~A., {et~al.} 2014, \apj, 788, 55,
  \dodoi{10.1088/0004-637X/788/1/55}

\bibitem[{{Bresolin} {et~al.}(2016){Bresolin}, {Kudritzki}, {Urbaneja},
  {Gieren}, {Ho}, \& {Pietrzy{\'n}ski}}]{bre16}
{Bresolin}, F., {Kudritzki}, R.-P., {Urbaneja}, M.~A., {et~al.} 2016, \apj,
  830, 64, \dodoi{10.3847/0004-637X/830/2/64}

\bibitem[{{Chatzikos} {et~al.}(2023){Chatzikos}, {Bianchi}, {Camilloni},
  {Chakraborty}, {Gunasekera}, {Guzm{\'a}n}, {Milby}, {Sarkar}, {Shaw}, {van
  Hoof}, \& {Ferland}}]{cha23}
{Chatzikos}, M., {Bianchi}, S., {Camilloni}, F., {et~al.} 2023, \rmxaa, 59,
  327, \dodoi{10.22201/ia.01851101p.2023.59.02.12}

\bibitem[{{Contini} {et~al.}(2004){Contini}, {Viegas}, \& {Prieto}}]{con04}
{Contini}, M., {Viegas}, S.~M., \& {Prieto}, M.~A. 2004, \mnras, 348, 1065,
  \dodoi{10.1111/j.1365-2966.2004.07430.x}

\bibitem[{{de Vaucouleurs} {et~al.}(1991){de Vaucouleurs}, {de Vaucouleurs},
  {Corwin}, {Buta}, {Paturel}, \& {Fouque}}]{vau91}
{de Vaucouleurs}, G., {de Vaucouleurs}, A., {Corwin}, Jr., H.~G., {et~al.}
  1991, {Third Reference Catalogue of Bright Galaxies}

\bibitem[{{Della Bruna} {et~al.}(2022){Della Bruna}, {Adamo}, {Amram},
  {Rosolowsky}, {Usher}, {Sirressi}, {Schruba}, {Emsellem}, {Leroy}, {Bik},
  {Blair}, {McLeod}, {{\"O}stlin}, {Renaud}, {Robert}, {Rousseau-Nepton}, \&
  {Smith}}]{del22}
{Della Bruna}, L., {Adamo}, A., {Amram}, P., {et~al.} 2022, \aap, 660, A77,
  \dodoi{10.1051/0004-6361/202142315}

\bibitem[{{Dopita} \& {Sutherland}(1996)}]{dop96}
{Dopita}, M.~A., \& {Sutherland}, R.~S. 1996, \apjs, 102, 161,
  \dodoi{10.1086/192255}

\bibitem[{{Dopita} \& {Sutherland}(2003)}]{dop03}
---. 2003, {Astrophysics of the diffuse universe},
  \dodoi{10.1007/978-3-662-05866-4}

\bibitem[{{Dopita} {et~al.}(2010){Dopita}, {Blair}, {Long}, {Mutchler},
  {Whitmore}, {Kuntz}, {Balick}, {Bond}, {Calzetti}, {Carollo}, {Disney},
  {Frogel}, {O'Connell}, {Hall}, {Holtzman}, {Kimble}, {MacKenty}, {McCarthy},
  {Paresce}, {Saha}, {Silk}, {Sirianni}, {Trauger}, {Walker}, {Windhorst}, \&
  {Young}}]{dop10}
{Dopita}, M.~A., {Blair}, W.~P., {Long}, K.~S., {et~al.} 2010, \apj, 710, 964,
  \dodoi{10.1088/0004-637X/710/2/964}

\bibitem[{{Fathi} {et~al.}(2008){Fathi}, {Beckman}, {Lundgren}, {Carignan},
  {Hernandez}, {Amram}, {Balard}, {Boulesteix}, {Gach}, {Knapen}, \&
  {Rela{\~n}o}}]{fah08}
{Fathi}, K., {Beckman}, J.~E., {Lundgren}, A.~A., {et~al.} 2008, \apjl, 675,
  L17, \dodoi{10.1086/527473}

\bibitem[{{Fern{\'a}ndez-Ontiveros} {et~al.}(2021){Fern{\'a}ndez-Ontiveros},
  {P{\'e}rez-Montero}, {V{\'\i}lchez}, {Amor{\'\i}n}, \& {Spinoglio}}]{fer21}
{Fern{\'a}ndez-Ontiveros}, J.~A., {P{\'e}rez-Montero}, E., {V{\'\i}lchez},
  J.~M., {Amor{\'\i}n}, R., \& {Spinoglio}, L. 2021, \aap, 652, A23,
  \dodoi{10.1051/0004-6361/202039716}

\bibitem[{{Feuillet} {et~al.}(2024){Feuillet}, {Kraemer}, {Mel{\'e}ndez},
  {Fischer}, {Schmitt}, {Reeves}, \& {Trindade Falc{\~a}o}}]{feu24}
{Feuillet}, L.~M., {Kraemer}, S., {Mel{\'e}ndez}, M.~B., {et~al.} 2024, arXiv
  e-prints, arXiv:2409.07665, \dodoi{10.48550/arXiv.2409.07665}

\bibitem[{{Fricke} {et~al.}(2001){Fricke}, {Izotov}, {Papaderos}, {Guseva}, \&
  {Thuan}}]{fri01}
{Fricke}, K.~J., {Izotov}, Y.~I., {Papaderos}, P., {Guseva}, N.~G., \& {Thuan},
  T.~X. 2001, \aj, 121, 169, \dodoi{10.1086/318016}

\bibitem[{{Gordon} {et~al.}(2021){Gordon}, {Misselt}, {Bouwman}, {Clayton},
  {Decleir}, {Hines}, {Pendleton}, {Rieke}, {Smith}, \& {Whittet}}]{gor21}
{Gordon}, K.~D., {Misselt}, K.~A., {Bouwman}, J., {et~al.} 2021, \apj, 916, 33,
  \dodoi{10.3847/1538-4357/ac00b7}

\bibitem[{{Groves} {et~al.}(2006){Groves}, {Dopita}, \& {Sutherland}}]{gro06}
{Groves}, B., {Dopita}, M., \& {Sutherland}, R. 2006, \aap, 458, 405,
  \dodoi{10.1051/0004-6361:20065097}

\bibitem[{{Gunasekera} {et~al.}(2023){Gunasekera}, {van Hoof}, {Chatzikos}, \&
  {Ferland}}]{gun23}
{Gunasekera}, C.~M., {van Hoof}, P. A.~M., {Chatzikos}, M., \& {Ferland}, G.~J.
  2023, Research Notes of the American Astronomical Society, 7, 246,
  \dodoi{10.3847/2515-5172/ad0e75}

\bibitem[{{Hao} {et~al.}(2009){Hao}, {Wu}, {Charmandaris}, {Spoon},
  {Bernard-Salas}, {Devost}, {Lebouteiller}, \& {Houck}}]{hao09}
{Hao}, L., {Wu}, Y., {Charmandaris}, V., {et~al.} 2009, \apj, 704, 1159,
  \dodoi{10.1088/0004-637X/704/2/1159}

\bibitem[{{Harris} {et~al.}(2001){Harris}, {Calzetti}, {Gallagher},
  {Conselice}, \& {Smith}}]{har01}
{Harris}, J., {Calzetti}, D., {Gallagher}, John~S., I., {Conselice}, C.~J., \&
  {Smith}, D.~A. 2001, \aj, 122, 3046, \dodoi{10.1086/324230}

\bibitem[{{Hermosa Mu{\~n}oz} {et~al.}(2024{\natexlab{a}}){Hermosa Mu{\~n}oz},
  {Alonso-Herrero}, {Pereira-Santaella}, {Garc{\'\i}a-Bernete},
  {Garc{\'\i}a-Burillo}, {Garc{\'\i}a-Lorenzo}, {Davies}, {Shimizu},
  {Esparza-Arredondo}, {Hicks}, {Haidar}, {Leist}, {L{\'o}pez-Rodr{\'\i}guez},
  {Ramos Almeida}, {Rosario}, {Zhang}, {Audibert}, {Bellocchi}, {Boorman},
  {Bunker}, {Combes}, {Campbell}, {D{\'\i}az-Santos}, {Fuller}, {Gandhi},
  {Gonz{\'a}lez-Mart{\'\i}n}, {H{\"o}nig}, {Imanishi}, {Izumi}, {Labiano},
  {Levenson}, {Packham}, {Ricci}, {Rigopoulou}, {Rouan}, {Stalevski},
  {Villar-Mart{\'\i}n}, \& {Ward}}]{her24}
{Hermosa Mu{\~n}oz}, L., {Alonso-Herrero}, A., {Pereira-Santaella}, M.,
  {et~al.} 2024{\natexlab{a}}, \aap, 690, A350,
  \dodoi{10.1051/0004-6361/202450262}

\bibitem[{{Hermosa Mu{\~n}oz} {et~al.}(2024{\natexlab{b}}){Hermosa Mu{\~n}oz},
  {Alonso-Herrero}, {Labiano}, {Guillard}, {Pantoni}, {Buiten}, {Dicken}, Baes,
  Boker, Colina, Donnan, Garcia-Bernete, Ostlin, van~der Werf, Ward, Brandl,
  Walter, Wright, Gudel, Henning, Lagage, \& Ray}]{her24b}
{Hermosa Mu{\~n}oz}, L., {Alonso-Herrero}, A., {Labiano}, A., {et~al.}
  2024{\natexlab{b}}, Astronomy \& Astrophysics,
  \dodoi{10.1051/0004-6361/202452437}

\bibitem[{{Hernandez} {et~al.}(2021){Hernandez}, {Aloisi}, {James}, {Kumari},
  {Berg}, {Adamo}, {Blair}, {Faucher-Gigu{\`e}re}, {Fox}, {Gurvich}, {Hafen},
  {Heckman}, {Lebouteiller}, {Long}, {Skillman}, {Tumlinson}, \&
  {Whitmore}}]{her21}
{Hernandez}, S., {Aloisi}, A., {James}, B.~L., {et~al.} 2021, \apj, 908, 226,
  \dodoi{10.3847/1538-4357/abd6c4}

\bibitem[{{Hernandez} {et~al.}(2023){Hernandez}, {Jones}, {Smith}, {Togi},
  {Aloisi}, {Blair}, {Hirschauer}, {Hunt}, {James}, {Kumari}, {Lebouteiller},
  {Mingozzi}, \& {Ramambason}}]{her23}
{Hernandez}, S., {Jones}, L., {Smith}, L.~J., {et~al.} 2023, \apj, 948, 124,
  \dodoi{10.3847/1538-4357/acc837}

\bibitem[{{Houghton} \& {Thatte}(2008)}]{hou08}
{Houghton}, R.~C.~W., \& {Thatte}, N. 2008, \mnras, 385, 1110,
  \dodoi{10.1111/j.1365-2966.2008.12893.x}

\bibitem[{{Izotov} {et~al.}(2001){Izotov}, {Chaffee}, \& {Schaerer}}]{izo01}
{Izotov}, Y.~I., {Chaffee}, F.~H., \& {Schaerer}, D. 2001, \aap, 378, L45,
  \dodoi{10.1051/0004-6361:20011265}

\bibitem[{{Izotov} {et~al.}(2004){Izotov}, {Noeske}, {Guseva}, {Papaderos},
  {Thuan}, \& {Fricke}}]{izo04}
{Izotov}, Y.~I., {Noeske}, K.~G., {Guseva}, N.~G., {et~al.} 2004, \aap, 415,
  L27, \dodoi{10.1051/0004-6361:20040006}

\bibitem[{{Izotov} \& {Thuan}(2000)}]{izo00}
{Izotov}, Y.~I., \& {Thuan}, T.~X. 2000, \nar, 44, 329,
  \dodoi{10.1016/S1387-6473(00)00044-0}

\bibitem[{{Izotov} {et~al.}(2021){Izotov}, {Thuan}, \& {Guseva}}]{izo21}
{Izotov}, Y.~I., {Thuan}, T.~X., \& {Guseva}, N.~G. 2021, \mnras, 508, 2556,
  \dodoi{10.1093/mnras/stab2798}

\bibitem[{{Jones} {et~al.}(2024){Jones}, {Hernandez}, {Smith}, {Togi},
  {Diaz-Santos}, {Aloisi}, {Blair}, {Hirschauer}, {Hunt}, {James}, {Kumari},
  {Lebouteiller}, {Mingozzi}, \& {Ramambason}}]{jon24}
{Jones}, L.~H., {Hernandez}, S., {Smith}, L.~J., {et~al.} 2024, arXiv e-prints,
  arXiv:2410.09020, \dodoi{10.48550/arXiv.2410.09020}

\bibitem[{{Kehrig} {et~al.}(2018){Kehrig}, {V{\'\i}lchez}, {Guerrero},
  {Iglesias-P{\'a}ramo}, {Hunt}, {Duarte-Puertas}, \& {Ramos-Larios}}]{keh18}
{Kehrig}, C., {V{\'\i}lchez}, J.~M., {Guerrero}, M.~A., {et~al.} 2018, \mnras,
  480, 1081, \dodoi{10.1093/mnras/sty1920}

\bibitem[{{Knapen} {et~al.}(2010){Knapen}, {Sharp}, {Ryder},
  {Falc{\'o}n-Barroso}, {Fathi}, \& {Guti{\'e}rrez}}]{kna10}
{Knapen}, J.~H., {Sharp}, R.~G., {Ryder}, S.~D., {et~al.} 2010, \mnras, 408,
  797, \dodoi{10.1111/j.1365-2966.2010.17180.x}

\bibitem[{{Kojima} {et~al.}(2021){Kojima}, {Ouchi}, {Rauch}, {Ono}, {Nakajima},
  {Isobe}, {Fujimoto}, {Harikane}, {Hashimoto}, {Hayashi}, {Komiyama},
  {Kusakabe}, {Kim}, {Lee}, {Mukae}, {Nagao}, {Onodera}, {Shibuya}, {Sugahara},
  {Umemura}, \& {Yabe}}]{koj21}
{Kojima}, T., {Ouchi}, M., {Rauch}, M., {et~al.} 2021, \apj, 913, 22,
  \dodoi{10.3847/1538-4357/abec3d}

\bibitem[{{Law} {et~al.}(2023){Law}, {E. Morrison}, {Argyriou}, {Patapis},
  {{\'A}lvarez-M{\'a}rquez}, {Labiano}, \& {Vandenbussche}}]{law23}
{Law}, D.~R., {E. Morrison}, J., {Argyriou}, I., {et~al.} 2023, \aj, 166, 45,
  \dodoi{10.3847/1538-3881/acdddc}

\bibitem[{{Leitherer} {et~al.}(1999){Leitherer}, {Schaerer}, {Goldader},
  {Delgado}, {Robert}, {Kune}, {de Mello}, {Devost}, \& {Heckman}}]{lei99}
{Leitherer}, C., {Schaerer}, D., {Goldader}, J.~D., {et~al.} 1999, \apjs, 123,
  3, \dodoi{10.1086/313233}

\bibitem[{{Long} {et~al.}(2022){Long}, {Blair}, {Winkler}, {Della Bruna},
  {Adamo}, {McLeod}, \& {Amram}}]{lon22}
{Long}, K.~S., {Blair}, W.~P., {Winkler}, P.~F., {et~al.} 2022, \apj, 929, 144,
  \dodoi{10.3847/1538-4357/ac5aa3}

\bibitem[{{Long} {et~al.}(2014){Long}, {Kuntz}, {Blair}, {Godfrey},
  {Plucinsky}, {Soria}, {Stockdale}, \& {Winkler}}]{lon14}
{Long}, K.~S., {Kuntz}, K.~D., {Blair}, W.~P., {et~al.} 2014, \apjs, 212, 21,
  \dodoi{10.1088/0067-0049/212/2/21}

\bibitem[{{Marshall} {et~al.}(2007){Marshall}, {Herter}, {Armus},
  {Charmandaris}, {Spoon}, {Bernard-Salas}, \& {Houck}}]{mar07}
{Marshall}, J.~A., {Herter}, T.~L., {Armus}, L., {et~al.} 2007, \apj, 670, 129,
  \dodoi{10.1086/521588}

\bibitem[{{Mart{\'\i}nez-Paredes} {et~al.}(2023){Mart{\'\i}nez-Paredes},
  {Bruzual}, {Morisset}, {Kim}, {Mel{\'e}ndez}, \& {Binette}}]{mar23}
{Mart{\'\i}nez-Paredes}, M., {Bruzual}, G., {Morisset}, C., {et~al.} 2023,
  \mnras, 525, 2916, \dodoi{10.1093/mnras/stad2447}

\bibitem[{{Mel{\'e}ndez} {et~al.}(2014){Mel{\'e}ndez}, {Heckman},
  {Mart{\'\i}nez-Paredes}, {Kraemer}, \& {Mendoza}}]{mel14}
{Mel{\'e}ndez}, M., {Heckman}, T.~M., {Mart{\'\i}nez-Paredes}, M., {Kraemer},
  S.~B., \& {Mendoza}, C. 2014, \mnras, 443, 1358,
  \dodoi{10.1093/mnras/stu1242}

\bibitem[{{Mel{\'e}ndez} {et~al.}(2011){Mel{\'e}ndez}, {Kraemer}, {Weaver}, \&
  {Mushotzky}}]{mel11}
{Mel{\'e}ndez}, M., {Kraemer}, S.~B., {Weaver}, K.~A., \& {Mushotzky}, R.~F.
  2011, \apj, 738, 6, \dodoi{10.1088/0004-637X/738/1/6}

\bibitem[{{Mel{\'e}ndez} {et~al.}(2008){Mel{\'e}ndez}, {Kraemer}, {Armentrout},
  {Deo}, {Crenshaw}, {Schmitt}, {Mushotzky}, {Tueller}, {Markwardt}, \&
  {Winter}}]{mel08}
{Mel{\'e}ndez}, M., {Kraemer}, S.~B., {Armentrout}, B.~K., {et~al.} 2008, \apj,
  682, 94, \dodoi{10.1086/588807}

\bibitem[{{Mingozzi} {et~al.}(2025){Mingozzi}, {Garcia Del Valle-Espinosa},
  {James}, {Rickards Vaught}, {Hayes}, {Amor{\'\i}n}, {Leitherer}, {Aloisi},
  {Hunt}, {Law}, {Richardson}, {Arellano-C{\'o}rdova}, {Berg}, {Chisholm},
  {Hernandez}, {Jones}, {Kumari}, {Martin}, {Ravindranath}, {Vallini}, \&
  {Xu}}]{min25}
{Mingozzi}, M., {Garcia Del Valle-Espinosa}, M., {James}, B.~L., {et~al.} 2025,
  arXiv e-prints, arXiv:2502.07662, \dodoi{10.48550/arXiv.2502.07662}

\bibitem[{{Mordini} {et~al.}(2021){Mordini}, {Spinoglio}, \&
  {Fern{\'a}ndez-Ontiveros}}]{mor21}
{Mordini}, S., {Spinoglio}, L., \& {Fern{\'a}ndez-Ontiveros}, J.~A. 2021, \aap,
  653, A36, \dodoi{10.1051/0004-6361/202140696}

\bibitem[{{Pakull} \& {Angebault}(1986)}]{pak86}
{Pakull}, M.~W., \& {Angebault}, L.~P. 1986, \nat, 322, 511,
  \dodoi{10.1038/322511a0}

\bibitem[{{Pereira-Santaella} {et~al.}(2010){Pereira-Santaella},
  {Diamond-Stanic}, {Alonso-Herrero}, \& {Rieke}}]{per10}
{Pereira-Santaella}, M., {Diamond-Stanic}, A.~M., {Alonso-Herrero}, A., \&
  {Rieke}, G.~H. 2010, \apj, 725, 2270, \dodoi{10.1088/0004-637X/725/2/2270}

\bibitem[{{Pereira-Santaella} {et~al.}(2022){Pereira-Santaella},
  {{\'A}lvarez-M{\'a}rquez}, {Garc{\'\i}a-Bernete}, {Labiano}, {Colina},
  {Alonso-Herrero}, {Bellocchi}, {Garc{\'\i}a-Burillo}, {H{\"o}nig}, {Ramos
  Almeida}, \& {Rosario}}]{per22}
{Pereira-Santaella}, M., {{\'A}lvarez-M{\'a}rquez}, J., {Garc{\'\i}a-Bernete},
  I., {et~al.} 2022, \aap, 665, L11, \dodoi{10.1051/0004-6361/202244725}

\bibitem[{{Richardson} {et~al.}(2022){Richardson}, {Simpson}, {Polimera},
  {Kannappan}, {Bellovary}, {Greene}, \& {Jenkins}}]{ric22}
{Richardson}, C.~T., {Simpson}, C., {Polimera}, M.~S., {et~al.} 2022, \apj,
  927, 165, \dodoi{10.3847/1538-4357/ac510c}

\bibitem[{{Russell} {et~al.}(2020){Russell}, {White}, {Long}, {Blair}, {Soria},
  \& {Winkler}}]{rus20}
{Russell}, T.~D., {White}, R.~L., {Long}, K.~S., {et~al.} 2020, \mnras, 495,
  479, \dodoi{10.1093/mnras/staa1177}

\bibitem[{{Rutledge} {et~al.}(2007){Rutledge}, {Bildsten}, {Brown},
  {Chakrabarty}, {Pavlov}, \& {Zavlin}}]{rut07}
{Rutledge}, R.~E., {Bildsten}, L., {Brown}, E.~F., {et~al.} 2007, \apj, 658,
  514, \dodoi{10.1086/510183}

\bibitem[{{Saha} {et~al.}(2006){Saha}, {Thim}, {Tammann}, {Reindl}, \&
  {Sandage}}]{sah06}
{Saha}, A., {Thim}, F., {Tammann}, G.~A., {Reindl}, B., \& {Sandage}, A. 2006,
  \apjs, 165, 108, \dodoi{10.1086/503800}

\bibitem[{{Satyapal} {et~al.}(2008){Satyapal}, {Vega}, {Dudik}, {Abel}, \&
  {Heckman}}]{sat08}
{Satyapal}, S., {Vega}, D., {Dudik}, R.~P., {Abel}, N.~P., \& {Heckman}, T.
  2008, \apj, 677, 926, \dodoi{10.1086/529014}

\bibitem[{{Satyapal} {et~al.}(2007){Satyapal}, {Vega}, {Heckman}, {O'Halloran},
  \& {Dudik}}]{sat07}
{Satyapal}, S., {Vega}, D., {Heckman}, T., {O'Halloran}, B., \& {Dudik}, R.
  2007, \apjl, 663, L9, \dodoi{10.1086/519995}

\bibitem[{{Schaerer}(1996)}]{sch96}
{Schaerer}, D. 1996, \apjl, 467, L17, \dodoi{10.1086/310193}

\bibitem[{{Schaerer} {et~al.}(2019){Schaerer}, {Fragos}, \& {Izotov}}]{sch19}
{Schaerer}, D., {Fragos}, T., \& {Izotov}, Y.~I. 2019, \aap, 622, L10,
  \dodoi{10.1051/0004-6361/201935005}

\bibitem[{{Schaerer} \& {Stasi{\'n}ska}(1999)}]{sch99}
{Schaerer}, D., \& {Stasi{\'n}ska}, G. 1999, \aap, 345, L17,
  \dodoi{10.48550/arXiv.astro-ph/9903430}

\bibitem[{{Senchyna} {et~al.}(2020){Senchyna}, {Stark}, {Mirocha}, {Reines},
  {Charlot}, {Jones}, \& {Mulchaey}}]{sen20}
{Senchyna}, P., {Stark}, D.~P., {Mirocha}, J., {et~al.} 2020, \mnras, 494, 941,
  \dodoi{10.1093/mnras/staa586}

\bibitem[{{Smith} {et~al.}(2002){Smith}, {Norris}, \& {Crowther}}]{smith02}
{Smith}, L.~J., {Norris}, R. P.~F., \& {Crowther}, P.~A. 2002, \mnras, 337,
  1309, \dodoi{10.1046/j.1365-8711.2002.06042.x}

\bibitem[{{Soria} \& {Wu}(2003)}]{sor03}
{Soria}, R., \& {Wu}, K. 2003, \aap, 410, 53,
  \dodoi{10.1051/0004-6361:20031074}

\bibitem[{{Spoon} {et~al.}(2022){Spoon}, {Hern{\'a}n-Caballero}, {Rupke},
  {Waters}, {Lebouteiller}, {Tielens}, {Loredo}, {Su}, \& {Viola}}]{spo22}
{Spoon}, H.~W.~W., {Hern{\'a}n-Caballero}, A., {Rupke}, D., {et~al.} 2022,
  \apjs, 259, 37, \dodoi{10.3847/1538-4365/ac4989}

\bibitem[{{Sturm} {et~al.}(2002){Sturm}, {Lutz}, {Verma}, {Netzer},
  {Sternberg}, {Moorwood}, {Oliva}, \& {Genzel}}]{stu02}
{Sturm}, E., {Lutz}, D., {Verma}, A., {et~al.} 2002, \aap, 393, 821,
  \dodoi{10.1051/0004-6361:20021043}

\bibitem[{{Sutherland} {et~al.}(1993){Sutherland}, {Bicknell}, \&
  {Dopita}}]{sut93}
{Sutherland}, R.~S., {Bicknell}, G.~V., \& {Dopita}, M.~A. 1993, \apj, 414,
  510, \dodoi{10.1086/173099}

\bibitem[{{Terlevich} {et~al.}(1995){Terlevich}, {Tenorio-Tagle}, {Rozyczka},
  {Franco}, \& {Melnick}}]{ter95}
{Terlevich}, R., {Tenorio-Tagle}, G., {Rozyczka}, M., {Franco}, J., \&
  {Melnick}, J. 1995, \mnras, 272, 198, \dodoi{10.1093/mnras/272.1.198}

\bibitem[{{Thatte} {et~al.}(2000){Thatte}, {Tecza}, \& {Genzel}}]{tha00}
{Thatte}, N., {Tecza}, M., \& {Genzel}, R. 2000, \aap, 364, L47,
  \dodoi{10.48550/arXiv.astro-ph/0009392}

\bibitem[{{Thuan} \& {Izotov}(2005)}]{thu05}
{Thuan}, T.~X., \& {Izotov}, Y.~I. 2005, \apjs, 161, 240,
  \dodoi{10.1086/491657}

\bibitem[{{Verma} {et~al.}(2003){Verma}, {Lutz}, {Sturm}, {Sternberg},
  {Genzel}, \& {Vacca}}]{ver03}
{Verma}, A., {Lutz}, D., {Sturm}, E., {et~al.} 2003, \aap, 403, 829,
  \dodoi{10.1051/0004-6361:20030408}

\bibitem[{{Weaver} {et~al.}(2010){Weaver}, {Mel{\'e}ndez}, {Mushotzky},
  {Kraemer}, {Engle}, {Malumuth}, {Tueller}, {Markwardt}, {Berghea}, {Dudik},
  {Winter}, \& {Armus}}]{wea10}
{Weaver}, K.~A., {Mel{\'e}ndez}, M., {Mushotzky}, R.~F., {et~al.} 2010, \apj,
  716, 1151, \dodoi{10.1088/0004-637X/716/2/1151}

\bibitem[{{Westmoquette} {et~al.}(2007){Westmoquette}, {Exter}, {Smith}, \&
  {Gallagher}}]{wes07}
{Westmoquette}, M.~S., {Exter}, K.~M., {Smith}, L.~J., \& {Gallagher}, J.~S.
  2007, \mnras, 381, 894, \dodoi{10.1111/j.1365-2966.2007.12346.x}

\bibitem[{{Yukita} {et~al.}(2016){Yukita}, {Hornschemeier}, {Lehmer}, {Ptak},
  {Wik}, {Zezas}, {Antoniou}, {Maccarone}, {Replicon}, {Tyler}, {Venters},
  {Argo}, {Bechtol}, {Boggs}, {Christensen}, {Craig}, {Hailey}, {Harrison},
  {Krivonos}, {Kuntz}, {Stern}, \& {Zhang}}]{yuk16}
{Yukita}, M., {Hornschemeier}, A.~E., {Lehmer}, B.~D., {et~al.} 2016, \apj,
  824, 107, \dodoi{10.3847/0004-637X/824/2/107}

\end{thebibliography}
\bibliographystyle{aasjournal}



\end{document}